\documentclass[1p]{elsarticle}

\usepackage{subcaption}
\usepackage{tikz}
\usepackage{pgfplots}
\usetikzlibrary{trees,positioning,arrows,decorations.pathreplacing, shapes,calc, external}
\usepackage{color}
\usepackage{amssymb}
\usepackage{amsmath, cases}
\usepackage{booktabs}
\usepackage{lineno,hyperref}
\usepackage{longtable}

\makeatletter
\def\ps@pprintTitle{%
	\let\@oddhead\@empty
	\let\@evenhead\@empty
	\def\@oddfoot{}%
	\let\@evenfoot\@oddfoot}
\makeatother

\newdefinition{rmk}{Remark}
\newproof{pf}{Proof}

\biboptions{authoryear}

\begin{document}

\begin{frontmatter}

\title{Predicting Item Popularity: Analysing Local Clustering Behaviour of Users} 

\author[mymainaddress]{J. Liebig} 
\author[mymainaddress]{A. Rao}
\address[mymainaddress]{School of Mathematical and Geospatial Sciences, RMIT University, Melbourne 3001, Australia}

\begin{abstract}
 Predicting the popularity of items in rating networks is an interesting but challenging problem. This is especially so when an item has first appeared and has received very few ratings. In this paper, we propose a novel approach to predicting the future popularity of new items in rating networks, defining a new bipartite clustering coefficient to predict the popularity of movies and stories in the MovieLens and Digg networks respectively. We show that the clustering behaviour of the first user who rates a new item gives insight into the future popularity of that item. Our method predicts, with a success rate of over 65\% for the MovieLens network and over 50\% for the Digg network, the future popularity of an item. This is a major improvement on current results. 
\end{abstract}


\end{frontmatter}

\section{Introduction}\label{sec:intro}
Websites like Amazon, TripAdvisor and MovieLens offer their users a means to rate a variety of different items. Users can decide whether they are interested in an item based on its previously received ratings. Websites collect these user ratings for many reasons including recommending items to their users and predicting future item ratings \citep{Resnick1994}. The latter is rather challenging. In particular, new items that have received very few ratings to date are hard to classify as being popular or unpopular in the future, due to the sparsity of information \citep{Schafer2007}. It has been suggested that a ranking of the users may aid in improving predictions of future item popularity \citep{Zeng2013}. In other words, the behaviour of some users may be adopted by others. In this paper, we predict the future popularity of new items by investigating the ego network of the user who first rated the new item. We use the ego's clustering behaviour to predict the number of ratings and the average rating the new item is likely to receive in the near future. We demonstrate our approach on the MovieLens network \citep{Grouplens2014}  that contains ratings of 10,681 movies by 71,567 different users and the Digg network \citep{Konect2014} that contains ratings of 3,553 stories by 139,409 users. Over 65\% of the time, we are able to predict the future popularity of new movies and over 50\% of the time we are able to predict the future popularity of new stories compared to a 30\% and 20\% prediction success respectively obtained by the current method \citep{Zeng2013}. 

Ranking the nodes of a network according to their importance or influence level is of high interest in many different areas and many ranking methods have been proposed \citep{Aral2012,Liu2013,Ren2014,Wei2013,Zhang2013}. A straightforward approach is to consider the node degrees. However, degree centrality is often unsuitable for identifying influential nodes, as a node with a few important neighbours may be more important than a node with a large number of unimportant neighbours \citep{Li2014}. Google's PageRank \citep{Page1999} takes this notion into account. This was improved by \citet{lu2011}, who proposed a parameter-free algorithm called LeaderRank that performs better than PageRank in many aspects. Another approach that uses k-shell decomposition and shows that the most important nodes of a network lie inside its core \citep{Kitsak2010}, was outperformed by a mixed degree decomposition procedure \citep{Zeng2013a} that considers the residual and exhausted degree of a node. \citet{Chen2013} point out that it is of advantage to design algorithms that are based on local information since networks are becoming larger. They propose a method that for the first time takes path diversity into account. Path diversity is very important in the identification of influential spreaders, since a node with many overlapping propagation paths may not spread, information for example, very efficiently.

All of the above approaches to finding influential nodes are designed for one-mode networks. In rating systems, on the other hand, users are not directly connected as all interactions take place between users and items. These systems are best represented by dynamic bipartite networks \citep{Zhou2007}. The nodes of a bipartite network can be partitioned into two disjoint sets, the primary and secondary node sets, such that edges only connect nodes from different sets \citep{Asratian1998}. Throughout this paper, primary nodes, the users, are denoted by $v_i$, whereas secondary nodes, the items, are denoted by $w_j$. 

\citet{Zeng2013} predict the popularity of items in three different bipartite networks. They define popularity by the increase in degree, meaning that an item with a high increase in ratings is considered as popular. It is reasonable to assume that popular items are rated more frequently, but there may be exceptions. Some items may receive a relatively high number of low ratings and consequently should not be considered as popular. To improve predictions \citet{Zeng2013} also consider user influence, where a user is considered to be influential if he shows high rating activity. However, high degree does not necessarily imply high influence \citep{Li2014}. Besides giving a large number of ratings, an influential user should also give a wide range of ratings. Although the approach by \citet{Zeng2013} works well for items that have been in the network for some time, with a success rate of 72\% for MovieLens, the fraction of new items correctly identified as popular in the future is small, approximately 30\%. The success rate for new items in the Digg network was 20\% and could be increased to 60\% by using the friendship network of the Digg users that is also available. Since for most bipartite rating networks, the underlying friendship network of users is not available we will not consider it in this study.

The main focus of this paper is to improve the prediction rate of new items. To do so, we propose a more sophisticated method, using the clustering coefficient. The clustering coefficient has been used in one-mode networks to identify influential nodes \citep{Chen2013a}. These influential nodes were found to spread information more quickly than those found via PageRank or LeaderRank. In their paper \citet{Chen2013a} investigate whether a node with lower clustering attracts more connections than a node with higher clustering. Their results clearly show that a node with a small clustering coefficient will build more connections in the future. Instead of focusing on the spreading abilities of nodes (users), our method improves predictions, especially of those items that are new in the system. Our approach only considers network topology and is suitable for data such as MovieLens, which does not record any particulars of the users, such as age or gender. For our method to work, the knowledge of single ratings, whether high or low, is also unnecessary. 

The rest of this paper is organised as follows: In Section \ref{sec:clustering}, we define a new clustering coefficient that suits the analysis of rating networks. Section \ref{sec:data} describes the analysed datasets and defines the term popularity in the context of rating networks. In Section \ref{sec:predictions}, we use our clustering coefficient as a tool to predict future popularity of new items, and present the results. Finally, we give conclusions and future work. 

\section{The Clustering Coefficient}\label{sec:clustering}
The clustering coefficient was originally defined for one-mode networks and, globally, measures the concentration of triangles \citep{Newman2010}. A bipartite network does not contain any cycles of odd length and hence is triangle free \citep{Asratian1998}. 

In \citet{Liebig2014}, we looked at previous definitions of the bipartite clustering coefficient \citep{Lind2005, Opsahl2013, Robins2004, Zhang2008} showing that it is important not only to consider triadic closure \citep{Opsahl2013} but to also distinguish between differently structured clusters. A cluster is a closed connection between three nodes of the same type and a 6-cycle is the only structure that could be considered a bipartite cluster \citep{Opsahl2013}.

\citet{Opsahl2013} calculates the bipartite clustering coefficient as follows:

\begin{equation}\label{eqn:clusteringOpsahl}
C^* = \frac{\textrm{number of closed 4-paths}}{\textrm{number of 4-paths}},
\end{equation}  

where a 4-path is a possible 6-cycle and a closed 4-path is a 6-cycle. Suppose there exists a 4-path $P=\{v_0, w_1, v_2, w_3, v_4\}$ at time $t_i$. At time $t_{i+1}$ a node $w_5$ is added to the network and connected to nodes $v_0$ and $v_4$, forming the 6-cycle $C^6=\{v_0, w_1, v_2, w_3, v_4, w_5, v_0\}$. 

\begin{figure}[h]
	\centering
	\begin{subfigure}[b]{0.4\textwidth}
		\centering
		\scalebox{1.0}{
			\begin{tikzpicture}[node distance=1cm, 
			c/.style={circle, thin, minimum width=1em, minimum height=1em, text centered, top color=gray!90!black, bottom color=gray!20!black, draw=black}, sq/.style={rectangle, thin, 		
				minimum width=1em, minimum height=1em, text centered, top color=gray!90!black, bottom color=gray!10!black, draw=black},
			sl/.style={-,>=stealth',auto,semithick,align = center,draw=black}, dl/.style={dotted,>=stealth',shorten >=1pt,auto,semithick,align = center,draw=black}]
			
			\node[c](c1){};
			\node[c, xshift = 2cm, yshift = 1cm](c2){};
			\node[c, xshift = 2cm, yshift = -1cm](c3){};
			\node[sq, xshift = 0.75cm, yshift = 1cm](r1){};
			\node[sq, xshift = 0.75cm, yshift = -1cm](r2){};
			\node[sq, xshift = 2.75cm, yshift = 0cm](r3){};
			
			\draw[sl] (c1) to node{}(r1);
			\draw[sl] (r1) to node{}(c2);
			\draw[sl] (c2) to node{}(r3);
			\draw[sl] (r3) to node{}(c3);
			\draw[sl] (c3) to node{}(r2);
			\draw[sl] (r2) to node{}(c1);
			\end{tikzpicture}
		}
		\caption{An induced 6-cycle.} \label{im:noIC6cycle}
	\end{subfigure}
	\qquad
	\begin{subfigure}[b]{0.4\textwidth}
		\centering
		\scalebox{1.0}{
			\begin{tikzpicture}[node distance=1cm, 
			c/.style={circle, thin, minimum width=1em, minimum height=1em, text centered, top color=gray!90!black, bottom color=gray!20!black, draw=black}, sq/.style={rectangle, thin, 		
				minimum width=1em, minimum height=1em, text centered, top color=gray!90!black, bottom color=gray!10!black, draw=black},
			sl/.style={-,>=stealth',auto,semithick,align = center,draw=black}, dl/.style={dotted,>=stealth',shorten >=1pt,auto,semithick,align = center,draw=black}]
			
			\node[c](c1){};
			\node[c, xshift = 2cm, yshift = 1cm](c2){};
			\node[c, xshift = 2cm, yshift = -1cm](c3){};
			\node[sq, xshift = 0.75cm, yshift = 1cm](r1){};
			\node[sq, xshift = 0.75cm, yshift = -1cm](r2){};
			\node[sq, xshift = 2.75cm, yshift = 0cm](r3){};
			
			\draw[sl] (c1) to node{}(r1);
			\draw[sl] (r1) to node{}(c2);
			\draw[sl] (c2) to node{}(r3);
			\draw[sl] (r3) to node{}(c3);
			\draw[sl] (c3) to node{}(r2);
			\draw[sl] (r2) to node{}(c1);
			
			\draw[dl] (r2) to node{}(c2);
			\end{tikzpicture}
		}
		\caption{A 6-cycle with one chord.} \label{im:oneIC6cycle}
	\end{subfigure}
	
	\begin{subfigure}[b]{0.4\textwidth}
		\centering
		\scalebox{1.0}{
			\begin{tikzpicture}[node distance=1cm, 
			c/.style={circle, thin, minimum width=1em, minimum height=1em, text centered, top color=gray!90!black, bottom color=gray!20!black, draw=black}, sq/.style={rectangle, thin, 		
				minimum width=1em, minimum height=1em, text centered, top color=gray!90!black, bottom color=gray!10!black, draw=black},
			sl/.style={-,>=stealth',auto,semithick,align = center,draw=black}, dl/.style={dotted,>=stealth',shorten >=1pt,auto,semithick,align = center,draw=black}]
			
			\node[c](c1){};
			\node[c, xshift = 2cm, yshift = 1cm](c2){};
			\node[c, xshift = 2cm, yshift = -1cm](c3){};
			\node[sq, xshift = 0.75cm, yshift = 1cm](r1){};
			\node[sq, xshift = 0.75cm, yshift = -1cm](r2){};
			\node[sq, xshift = 2.75cm, yshift = 0cm](r3){};
			
			\draw[sl] (c1) to node{}(r1);
			\draw[sl] (r1) to node{}(c2);
			\draw[sl] (c2) to node{}(r3);
			\draw[sl] (r3) to node{}(c3);
			\draw[sl] (c3) to node{}(r2);
			\draw[sl] (r2) to node{}(c1);	
			
			\draw[dl] (r2) to node{}(c2);
			\draw[dl] (r1) to node{}(c3);
			\end{tikzpicture}
		}
		\caption{A 6-cycle with two chords. }\label{im:twoIC6cycle}
	\end{subfigure}
	\qquad
	\begin{subfigure}[b]{0.4\textwidth}
		\centering
		\scalebox{1.0}{
			\begin{tikzpicture}[node distance=1cm, 
			c/.style={circle, thin, minimum width=1em, minimum height=1em, text centered, top color=gray!90!black, bottom color=gray!20!black, draw=black}, sq/.style={rectangle, thin, 		
				minimum width=1em, minimum height=1em, text centered, top color=gray!90!black, bottom color=gray!10!black, draw=black},
			sl/.style={-,>=stealth',auto,semithick,align = center,draw=black}, dl/.style={dotted,>=stealth',shorten >=1pt,auto,semithick,align = center,draw=black}]
			
			\node[c](c1){};
			\node[c, xshift = 2cm, yshift = 1cm](c2){};
			\node[c, xshift = 2cm, yshift = -1cm](c3){};
			\node[sq, xshift = 0.75cm, yshift = 1cm](r1){};
			\node[sq, xshift = 0.75cm, yshift = -1cm](r2){};
			\node[sq, xshift = 2.75cm, yshift = 0cm](r3){};	
			
			\draw[sl] (c1) to node{}(r1);
			\draw[sl] (r1) to node{}(c2);
			\draw[sl] (c2) to node{}(r3);
			\draw[sl] (r3) to node{}(c3);
			\draw[sl] (c3) to node{}(r2);
			\draw[sl] (r2) to node{}(c1);	
			
			\draw[dl] (c1) to node{}(r3);
			\draw[dl] (r2) to node{}(c2);
			\draw[dl] (r1) to node{}(c3);
			\end{tikzpicture}
		}
		\caption{A 6-cycle with three chords.} \label{im:threeIC6cycle}
	\end{subfigure}
	\caption{There are four different possible types of 6-cycles in a bipartite network: (a) an induced 6-cycle, (b) a 6-cycle with one chord, (c) a 6-cycle with two chords and (d) a 6-cycle with three chords. Chords are represented by dashed lines. \label{im:innerConnections6cycle}}
\end{figure}
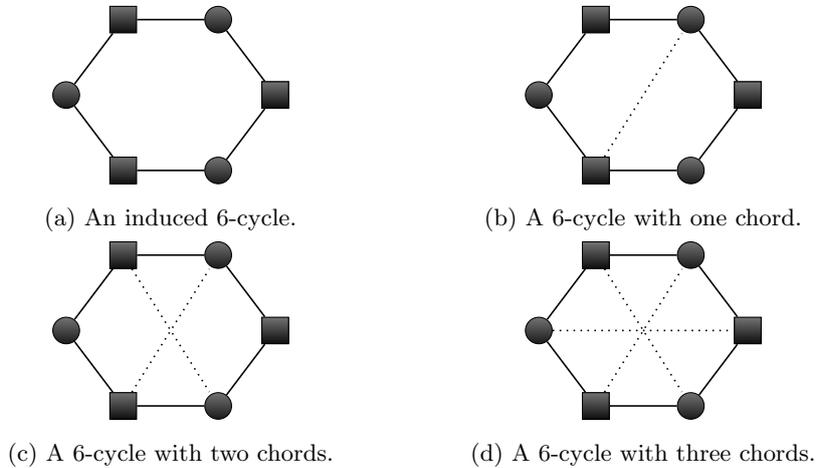

A 6-cycle in a bipartite network may have a maximum of three chords, resulting in four different types of cycles of length 6, see Fig. \ref{im:innerConnections6cycle}. A chord is an edge that connects two nodes of a cycle but is not itself part of the cycle. A cycle without chords is called an induced cycle \citep{Diestel2005}. \citet{Opsahl2013} does not distinguish between the different structures, leading to an over count of 6-cycles \citep{Liebig2014}.  

Further, Eq. \eqref{eqn:clusteringOpsahl} indirectly assumes that the network in question is formed in a certain way: At each time step $t_i$ one secondary node is added to the network and connected to an arbitrary number of primary nodes. In reality, only some networks develop in this manner, for instance, a network that models the attendance of people at events. In such a network, at each time step $t_i$ an event is added to the network and people make connections to this new event at the same time by attending it. Since a person can attend an event only at the time it takes place, people can form a connection to a particular event only at a particular point in time. We studied such types of networks in \citep{Liebig2014}.

Rating networks, on the other hand, develop very differently. Users can form connections to items at any point in time. In other words, at each time step $t_i$ one edge is added to the network and the assumption made by Eq. \eqref{eqn:clusteringOpsahl} is violated, as there is no necessity for a secondary node to be added at the same time, in order for a 6-cycle to be formed. Rating networks build a very different type of bipartite network and need to be analysed differently.     

\begin{figure}[h]
	\centering
	\begin{subfigure}[t]{0.4\textwidth}
		\centering
		\scalebox{0.4}{
			\begin{tikzpicture}[node distance=1cm, 
			c/.style={circle, thin, minimum width=1em, minimum height=1em, text centered, top color=gray!90!black, bottom color=gray!20!black, draw=black}, sq/.style={rectangle, thin, 		
				minimum width=1em, minimum height=1em, text centered, top color=gray!90!black, bottom color=gray!10!black, draw=black},
			sl/.style={-,>=stealth',auto,semithick,align = center,draw=black}, dl/.style={dotted,>=stealth',shorten >=1pt,auto,semithick,align = center,draw=black}]
			
			\node[c, label=left:{$v_0$}](v0){};
			\node[c, label=right:{$v_1$}, xshift = 2cm, yshift = 1cm](v1){};
			\node[c, label=right:{$v_2$}, xshift = 2cm, yshift = -1cm](v2){};
			\node[sq, label=left:{$w_0$}, xshift = 0.75cm, yshift = 1cm](n0){};
			\node[sq, label=left:{$w_2$},xshift = 0.75cm, yshift = -1cm](n2){};
			\node[sq, label=right:{$w_1$}, xshift = 2.75cm, yshift = 0cm](n1){};	
			
			\draw[sl] (v0) to node{}(n0);
			\draw[sl] (v0) to node{}(n2);
			\draw[sl] (v1) to node{}(n0);
			\draw[sl] (v1) to node{}(n1);
			\draw[sl] (v2) to node{}(n2);

			\node[c, label=left:{$v_0$}, xshift = 7.5cm, yshift = 0cm](v0){};
			\node[c, label=right:{$v_1$}, xshift = 9.5cm, yshift = 1cm](v1){};
			\node[c, label=right:{$v_2$}, xshift = 9.5cm, yshift = -1cm](v2){};
			\node[sq, label=left:{$w_0$}, xshift = 8.25cm, yshift = 1cm](n0){};
			\node[sq, label=left:{$w_2$},xshift = 8.25cm, yshift = -1cm](n2){};
			\node[sq, label=right:{$w_1$}, xshift = 10.25cm, yshift = 0cm](n1){};	
			
			\draw[sl] (v0) to node{}(n0);
			\draw[sl] (v0) to node{}(n2);
			\draw[sl] (v1) to node{}(n0);
			\draw[sl] (v1) to node{}(n1);
			\draw[sl] (v2) to node{}(n1);
			\draw[sl] (v2) to node{}(n2);
			
			\draw[thick,->] (4,0) -- (6.25,0);
			\end{tikzpicture}
		}
		\caption{We call the origin of an induced 6-cycle an unconnected 5-path.} \label{im:originInduced6cycle}
	\end{subfigure}
	\quad
	\begin{subfigure}[t]{0.4\textwidth}
		\centering
		\scalebox{0.4}{
			\begin{tikzpicture}[node distance=1cm, 
			c/.style={circle, thin, minimum width=1em, minimum height=1em, text centered, top color=gray!90!black, bottom color=gray!20!black, draw=black}, sq/.style={rectangle, thin, 		
				minimum width=1em, minimum height=1em, text centered, top color=gray!90!black, bottom color=gray!10!black, draw=black},
			sl/.style={-,>=stealth',auto,semithick,align = center,draw=black}, dl/.style={dotted,>=stealth',shorten >=1pt,auto,semithick,align = center,draw=black}]
			
			\node[c, label=left:{$v_0$}, xshift = 4.75cm, yshift = 0cm](c1){};
			\node[c, label=right:{$v_1$}, xshift = 6.75cm, yshift = 1cm](c2){};
			\node[c, label=right:{$v_2$}, xshift = 6.75cm, yshift = -1cm](c3){};
			\node[sq, label=left:{$w_0$}, xshift = 5.5cm, yshift = 1cm](r1){};
			\node[sq, label=left:{$w_2$},xshift = 5.5cm, yshift = -1cm](r2){};
			\node[sq, label=right:{$w_1$}, xshift = 7.5cm, yshift = 0cm](r3){};	
			
			\draw[sl] (c1) to node{}(r1);
			\draw[sl] (r1) to node{}(c2);
			\draw[sl] (c2) to node{}(r3);
			\draw[sl] (r3) to node{}(c3);
			\draw[sl] (c3) to node{}(r2);
			\draw[sl] (r2) to node{}(c1);
			
			\draw[dl] (c1) to node{}(r3);
			
			\node[c, label=left:{$v_0$}, xshift = 0cm, yshift = 2cm](c4){};
			\node[c, label=right:{$v_1$}, xshift = 2cm, yshift = 3cm](c5){};
			\node[c, label=right:{$v_2$}, xshift = 2cm, yshift = 1cm](c6){};
			\node[sq, label=left:{$w_0$}, xshift = 0.75cm, yshift = 3cm](r4){};
			\node[sq, label=left:{$w_2$},xshift = 0.75cm, yshift = 1cm](r5){};
			\node[sq, label=right:{$w_1$}, xshift = 2.75cm, yshift = 2cm](r6){};
			
			\draw[sl] (c4) to node{}(r4);
			\draw[sl] (r4) to node{}(c5);
			\draw[sl] (c5) to node{}(r6);
			\draw[sl] (c6) to node{}(r5);
			\draw[sl] (r5) to node{}(c4);	
			
			\draw[dl] (c4) to node{}(r6);
			
			\node[c, label=left:{$v_0$}, xshift = 9.5cm, yshift = 2cm](c4){};
			\node[c, label=right:{$v_1$}, xshift = 11.5cm, yshift = 3cm](c5){};
			\node[c, label=right:{$v_2$}, xshift = 11.5cm, yshift = 1cm](c6){};
			\node[sq, label=left:{$w_0$}, xshift = 10.25cm, yshift = 3cm](r4){};
			\node[sq, label=left:{$w_2$},xshift = 10.25cm, yshift = 1cm](r5){};
			\node[sq, label=right:{$w_1$}, xshift = 12.25cm, yshift = 2cm](r6){};
			
			\draw[sl] (c4) to node{}(r4);
			\draw[sl] (r6) to node{}(c6);
			\draw[sl] (c5) to node{}(r6);
			\draw[sl] (c6) to node{}(r5);
			\draw[sl] (r5) to node{}(c4);	
			
			\draw[dl] (c4) to node{}(r6);
			
			\draw[thick,->] (3,1.5) -- (4.25,1);
			\draw[thick,->] (9.25,1.5) -- (7.75,1);
			\draw[thick,->] (3,-1.5) -- (4.25,-1);
			\draw[thick,->] (9.25,-1.5) -- (7.75,-1);
			
			\node[c, label=left:{$v_0$}, xshift = 0cm, yshift = -2cm](c1){};
			\node[c, label=right:{$v_1$}, xshift = 2cm, yshift = -1cm](c2){};
			\node[c, label=right:{$v_2$}, xshift = 2cm, yshift = -3cm](c3){};
			\node[sq, label=left:{$w_0$}, xshift = 0.75cm, yshift = -1cm](r1){};
			\node[sq, label=left:{$w_2$},xshift = 0.75cm, yshift = -3cm](r2){};
			\node[sq, label=right:{$w_1$}, xshift = 2.75cm, yshift = -2cm](r3){};	
			
			\draw[sl] (c1) to node{}(r1);
			\draw[sl] (r1) to node{}(c2);
			\draw[sl] (c2) to node{}(r3);
			\draw[sl] (c3) to node{}(r2);
			\draw[sl] (r2) to node{}(c1);
			\draw[sl] (r3) to node{}(c3);

			\node[c, label=left:{$v_0$}, xshift = 9.5cm, yshift = -2cm](c1){};
			\node[c, label=right:{$v_1$}, xshift = 11.5cm, yshift = -1cm](c2){};
			\node[c, label=right:{$v_2$}, xshift = 11.5cm, yshift = -3cm](c3){};
			\node[sq, label=left:{$w_0$}, xshift = 10.25cm, yshift = -1cm](r1){};
			\node[sq, label=left:{$w_2$},xshift = 10.25cm, yshift = -3cm](r2){};
			\node[sq, label=right:{$w_1$}, xshift = 12.25cm, yshift = -2cm](r3){};	
			
			\draw[sl] (c1) to node{}(r1);
			\draw[sl] (r1) to node{}(c2);
			\draw[sl] (c2) to node{}(r3);
			\draw[sl] (c3) to node{}(r2);
			\draw[sl] (r3) to node{}(c3);
			
			\draw[dl] (c1) to node{}(r3);
			
			\end{tikzpicture}
		}
		\caption{An induced 6-cycle may form a 6-cycle with one chord at the next time step. We call a 5-path that contains an extra edge between two of its nodes that does not belong to the path, a connected 5-path. Any of the three different connected 5-paths above may also form a 6-cycle with one chord.}  \label{im:originOneChord6cycle}
	\end{subfigure}
	
	\begin{subfigure}[t]{0.4\textwidth}
		\centering
		\scalebox{0.4}{
			\begin{tikzpicture}[node distance=1cm, 
			c/.style={circle, thin, minimum width=1em, minimum height=1em, text centered, top color=gray!90!black, bottom color=gray!20!black, draw=black}, sq/.style={rectangle, thin, 		
				minimum width=1em, minimum height=1em, text centered, top color=gray!90!black, bottom color=gray!10!black, draw=black},
			sl/.style={-,>=stealth',auto,semithick,align = center,draw=black}, dl/.style={dotted,>=stealth',shorten >=1pt,auto,semithick,align = center,draw=black}]
			
			\node[c, label=left:{$v_0$}, xshift = 4.75cm, yshift = 0cm](c1){};
			\node[c, label=right:{$v_1$}, xshift = 6.75cm, yshift = 1cm](c2){};
			\node[c, label=right:{$v_2$}, xshift = 6.75cm, yshift = -1cm](c3){};
			\node[sq, label=left:{$w_0$}, xshift = 5.5cm, yshift = 1cm](r1){};
			\node[sq, label=left:{$w_2$},xshift = 5.5cm, yshift = -1cm](r2){};
			\node[sq, label=right:{$w_1$}, xshift = 7.5cm, yshift = 0cm](r3){};	
			
			\draw[sl] (c1) to node{}(r1);
			\draw[sl] (r1) to node{}(c2);
			\draw[sl] (c2) to node{}(r3);
			\draw[sl] (r3) to node{}(c3);
			\draw[sl] (c3) to node{}(r2);
			\draw[sl] (r2) to node{}(c1);
			
			\draw[dl] (c1) to node{}(r3);
			\draw[dl] (c2) to node{}(r2);
			
			\node[c, label=left:{$v_0$}, xshift = 0cm, yshift = 2cm](c4){};
			\node[c, label=right:{$v_1$}, xshift = 2cm, yshift = 3cm](c5){};
			\node[c, label=right:{$v_2$}, xshift = 2cm, yshift = 1cm](c6){};
			\node[sq, label=left:{$w_0$}, xshift = 0.75cm, yshift = 3cm](r4){};
			\node[sq, label=left:{$w_2$},xshift = 0.75cm, yshift = 1cm](r5){};
			\node[sq, label=right:{$w_1$}, xshift = 2.75cm, yshift = 2cm](r6){};
			
			\draw[sl] (c4) to node{}(r4);
			\draw[sl] (r4) to node{}(c5);
			\draw[sl] (c5) to node{}(r6);
			\draw[sl] (c6) to node{}(r5);
			\draw[sl] (r5) to node{}(c4);	
			
			\draw[dl] (c4) to node{}(r6);
			\draw[dl] (c5) to node{}(r5);
			
			\node[c, label=left:{$v_0$}, xshift = 9.5cm, yshift = 2cm](c4){};
			\node[c, label=right:{$v_1$}, xshift = 11.5cm, yshift = 3cm](c5){};
			\node[c, label=right:{$v_2$}, xshift = 11.5cm, yshift = 1cm](c6){};
			\node[sq, label=left:{$w_0$}, xshift = 10.25cm, yshift = 3cm](r4){};
			\node[sq, label=left:{$w_2$},xshift = 10.25cm, yshift = 1cm](r5){};
			\node[sq, label=right:{$w_1$}, xshift = 12.25cm, yshift = 2cm](r6){};
			
			\draw[sl] (c4) to node{}(r4);
			\draw[sl] (r6) to node{}(c6);
			\draw[sl] (c5) to node{}(r6);
			\draw[sl] (c6) to node{}(r5);
			\draw[sl] (r5) to node{}(c4);	
			
			\draw[dl] (c4) to node{}(r6);
			\draw[dl] (c5) to node{}(r5);
			
			\draw[thick,->] (3,1.5) -- (4.25,1);
			\draw[thick,->] (9.25,1.5) -- (7.75,1);
			\draw[thick,->] (3,-1.5) -- (4.25,-1);
			
			\node[c, label=left:{$v_0$}, xshift = 0cm, yshift = -2cm](c1){};
			\node[c, label=right:{$v_1$}, xshift = 2cm, yshift = -1cm](c2){};
			\node[c, label=right:{$v_2$}, xshift = 2cm, yshift = -3cm](c3){};
			\node[sq, label=left:{$w_0$}, xshift = 0.75cm, yshift = -1cm](r1){};
			\node[sq, label=left:{$w_2$},xshift = 0.75cm, yshift = -3cm](r2){};
			\node[sq, label=right:{$w_1$}, xshift = 2.75cm, yshift = -2cm](r3){};	
			
			\draw[sl] (c1) to node{}(r1);
			\draw[sl] (r1) to node{}(c2);
			\draw[sl] (c2) to node{}(r3);
			\draw[sl] (c3) to node{}(r2);
			\draw[sl] (r2) to node{}(c1);
			\draw[sl] (r3) to node{}(c3);
			
			\draw[dl] (c1) to node{}(r3);
			
			\end{tikzpicture}
		}
		\caption{A 6-cycle with one chord may form a 6-cycle with two chords at the next time step. We call a 5-path that contains two extra edges between its nodes that do not belong to the path, a completely connected 5-path. Any of the two different completely connected 5-paths above may also form a 6-cycle with two chords.} \label{im:originTwoChord6cycle}
	\end{subfigure}
	\quad
	\begin{subfigure}[t]{0.4\textwidth}
		\centering
		\scalebox{0.4}{
			\begin{tikzpicture}[node distance=1cm, 
			c/.style={circle, thin, minimum width=1em, minimum height=1em, text centered, top color=gray!90!black, bottom color=gray!20!black, draw=black}, sq/.style={rectangle, thin, 		
				minimum width=1em, minimum height=1em, text centered, top color=gray!90!black, bottom color=gray!10!black, draw=black},
			sl/.style={-,>=stealth',auto,semithick,align = center,draw=black}, dl/.style={dotted,>=stealth',shorten >=1pt,auto,semithick,align = center,draw=black}]
			
			\node[c, label=left:{$v_0$}](v0){};
			\node[c, label=right:{$v_1$}, xshift = 2cm, yshift = 1cm](v1){};
			\node[c, label=right:{$v_2$}, xshift = 2cm, yshift = -1cm](v2){};
			\node[sq, label=left:{$w_0$}, xshift = 0.75cm, yshift = 1cm](n0){};
			\node[sq, label=left:{$w_2$},xshift = 0.75cm, yshift = -1cm](n2){};
			\node[sq, label=right:{$w_1$}, xshift = 2.75cm, yshift = 0cm](n1){};	
			
			\draw[sl] (v0) to node{}(n0);
			\draw[sl] (v0) to node{}(n2);
			\draw[sl] (v1) to node{}(n0);
			\draw[sl] (v1) to node{}(n1);
			\draw[sl] (v2) to node{}(n2);
			\draw[sl] (v2) to node{}(n1);
			
			\draw[dl] (v1) to node{}(n2);
			\draw[dl] (v0) to node{}(n1);

			\node[c, label=left:{$v_0$}, xshift = 7.5cm, yshift = 0cm](v0){};
			\node[c, label=right:{$v_1$}, xshift = 9.5cm, yshift = 1cm](v1){};
			\node[c, label=right:{$v_2$}, xshift = 9.5cm, yshift = -1cm](v2){};
			\node[sq, label=left:{$w_0$}, xshift = 8.25cm, yshift = 1cm](n0){};
			\node[sq, label=left:{$w_2$},xshift = 8.25cm, yshift = -1cm](n2){};
			\node[sq, label=right:{$w_1$}, xshift = 10.25cm, yshift = 0cm](n1){};	
			
			\draw[sl] (v0) to node{}(n0);
			\draw[sl] (v0) to node{}(n2);
			\draw[sl] (v1) to node{}(n0);
			\draw[sl] (v1) to node{}(n1);
			\draw[sl] (v2) to node{}(n1);
			\draw[sl] (v2) to node{}(n2);
			
			\draw[dl] (v1) to node{}(n2);
			\draw[dl] (v0) to node{}(n1);
			\draw[dl] (v2) to node{}(n0);
			
			\draw[thick,->] (4,0) -- (6.25,0);
			\end{tikzpicture}
		}
		\caption{A 6-cycle with two chords may form a 6-cycle with three chords at the next time step.}\label{im:originThreeChord6cycle}
	\end{subfigure}
	\caption{All possibilities by which the different 6-cycles may be formed in a rating network.} \label{im:formation6cyclesType2}
\end{figure}
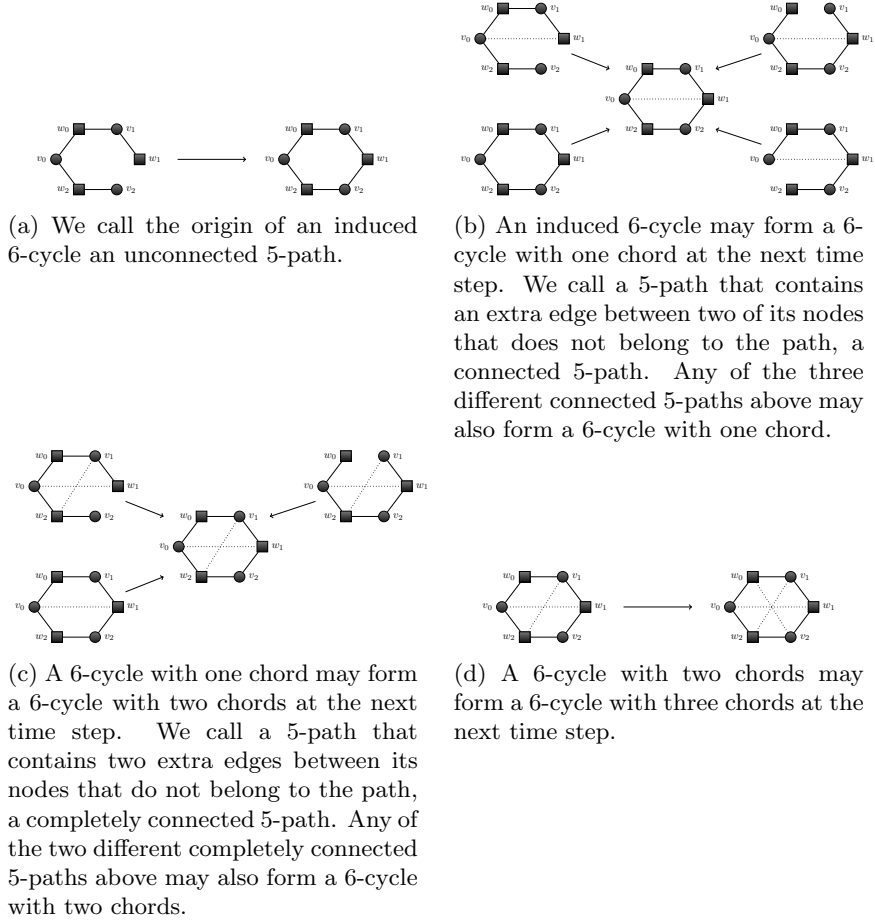

To find a clustering coefficient that agrees with the development, over time, of rating networks, we examine the formation of the different 6-cycles, see Fig. \ref{im:innerConnections6cycle}, in rating networks. In the following we make use of induced subgraphs. A subgraph of a network is induced if it contains all the edges that connect the nodes of the subgraph in the original network \citep{Diestel2005}. The different ways in which 6-cycles in rating networks may be formed are depicted in Fig. \ref{im:formation6cyclesType2}.  

Equations \eqref{eqn:cci0} - \eqref{eqn:cci3} give four novel clustering coefficients $icc_{(k)}$, one for each type of 6-cycle, that suit the analysis of rating networks.

The induced clustering coefficient is given by:

\begin{equation}\label{eqn:cci0}
icc_{(0)} = \frac{6\sigma_{(0)}}{6\sigma_{(0)} + \kappa_{(0)}},
\end{equation}

where $\sigma_{(0)}$ is the number of induced 6-cycles and $\kappa_{(0)}$ is the number of unconnected 5-paths. The induced clustering coefficient $icc_{(0)}$ measures the proportion of induced 6-cycles to all unconnected paths of length 5 (Fig. \ref{im:originInduced6cycle}). The number of induced 6-cycles is multiplied by 6, as each induced 6-cycle contains six unconnected 5-paths.

The one chord clustering coefficient is given by:

\begin{equation}\label{eqn:cci1}
icc_{(1)} = \frac{7\sigma_{(1)}}{7\sigma_{(1)} + \sigma_{(0)} + \kappa_{(1)}},
\end{equation}

where $\sigma_{(1)}$ is the number of induced 6-cycles with one chord and $\kappa_{(1)}$ is the number of connected 5-paths. The one chord clustering coefficient $icc_{(1)}$ measures the proportion of 6-cycles with one chord with respect to its origins (Fig \ref{im:originOneChord6cycle}). The number of 6-cycles with one chord is multiplied by 7, as each contains two of each of the connected 5-paths, shown in Fig. \ref{im:originOneChord6cycle}, and one 6-cycle without any chords. 

The two chord clustering coefficient is given by:

\begin{equation}\label{eqn:cci2}
icc_{(2)} = \frac{4\sigma_{(2)}}{4\sigma_{(2)} + \sigma_{(1)} + \kappa_{(2)}},
\end{equation}

where $\sigma_{(2)}$ is the number of induced 6-cycles with two chords and $\kappa_{(2)}$ is the number of completely connected 5-paths. The two chord clustering coefficient $icc_{(2)}$ measures the proportion of 6-cycles with two chords with respect to its origins (Fig. \ref{im:originTwoChord6cycle}). The number of 6-cycles with two chords is multiplied by 4, as it contains one of each of the completely connected 5-paths, shown in Fig. \ref{im:originTwoChord6cycle}, and two 6-cycles with one chord. 

The three chord clustering coefficient is given by:

\begin{equation}\label{eqn:cci3}
icc_{(3)} = \frac{3\sigma_{(3)}}{3\sigma_{(3)} + \sigma_{(2)}},
\end{equation}

where $\sigma_{(3)}$ is the number of induced 6-cycles with three chords. The three chord clustering coefficient $icc_{(3)}$ measures the proportion of 6-cycles with respect to its origin (Fig. \ref{im:originThreeChord6cycle}). The number of 6-cycles with three chords is multiplied by 3, as it contains three 6-cycles with two chords.  

In a one-mode network, the local clustering coefficient of a node $v_i$ is calculated by dividing the number of closed 2-paths that are centred at node $v_i$ by the total number of 2-paths that are centred at node $v_i$. In rating networks, most clusters are formed from 5-paths (see Fig. \ref{im:formation6cyclesType2}). As a path of odd length can never be centred at a node, we consider all paths that involve node $v_i$ in order to calculate $v_i$'s clustering coefficient. The local clustering coefficients are denoted $icc_{(i,k)}$.

\section{The Data}\label{sec:data}
To test our prediction method, we have chosen two different datasets.

\subsection{The MovieLens Dataset}

The MovieLens data \citep{Grouplens2014} was collected by the University of Minnesota and contains 10,000,054 movie ratings that range between 1 and 5, with 5 being the best possible rating. Starting in January 1995, 71,567 different users rated 10,681 movies over a period of 14 years. Every user was assigned a unique id but no additional information about the users is known. A network is formed taking the users as the primary nodes and the movies as the secondary nodes. Each rating of a movie by a user is represented by an edge that links the user to the movie. Every edge is associated with a time stamp that corresponds to the time the rating was made. 

\subsection{The Digg Dataset}
Digg (http://digg.com/) is a website that presents news stories and allows users to vote for them. This bipartite network contains 3,018,197 votes cast by 139,409 users. A total of 3,553 stories were rated over a period of one month in 2009. Unlike the MovieLens network, edges are not associated with a rating. An edge between a user and a story indicates that the user liked the story. The data was obtained from \citet{Konect2014}.

\section{Predicting Popularity}\label{sec:predictions}
In this paper, we predict the popularity of newly released items using a novel approach. Firstly, we directly analyse the bipartite network, without using its one-mode projection in order to overcome information loss \citep{Conaldi2012, Vogt2010, Zhou2007}. Secondly, we use our newly defined clustering coefficient as a  tool to make accurate item popularity predictions.  

\subsection{What does popularity mean?}
Before making predictions about an item's future popularity, it is important to clearly define the meaning of popular and unpopular. 

In previous research, \citep{Zeng2013}, an item was considered popular if its degree increased rapidly over a certain period of time. Intuitively, a popular movie, for example, is watched and rated more often than an unpopular movie, however, there are exceptions. For instance, some movies may receive a relatively high number of low ratings and certainly should not be considered as popular. 

In contrast to the preferential attachment model \citep{Barabasi1999, Price1976} that predicts that nodes with a high degree are more likely to increase their degree than nodes with a low degree, in rating networks it is generally the case that the interest in an item, such as a movie, decays over time \citep{Medo2011}. The MovieLens data shows that, although movies are rated over a longer period of time, in many cases the ratings made within one month of the movie's release determine its final average rating, see Fig. S1 in the supplementary material. In case of the Digg dataset an item is frequently rated within 48 hours after the initial rating. After this period the interest in the item decays very quickly, see Fig. S2 in the supplementary material. Hence there is a need for making good early predictions.

We calculate the average number of ratings that items receive during this critical period. We define the critical period as the time period that most affects the average rating of an item. In case of the Digg network, where edges are not associated with a rating, the critical period is the time span in which stories are most frequently rated, i.e. within 48 hours of the initial rating. An item is considered as popular if it receives a higher number of ratings than the average item and obtains a high average rating. In the MovieLens dataset for instance, a movie received 29 ratings on average during the first month. Hence a movie is considered as popular if it receives 29 ratings or more during the first month and obtains an average rating greater or equal to 4.

\begin{figure}[h]
	\centering
	\includegraphics[width=0.8\textwidth]{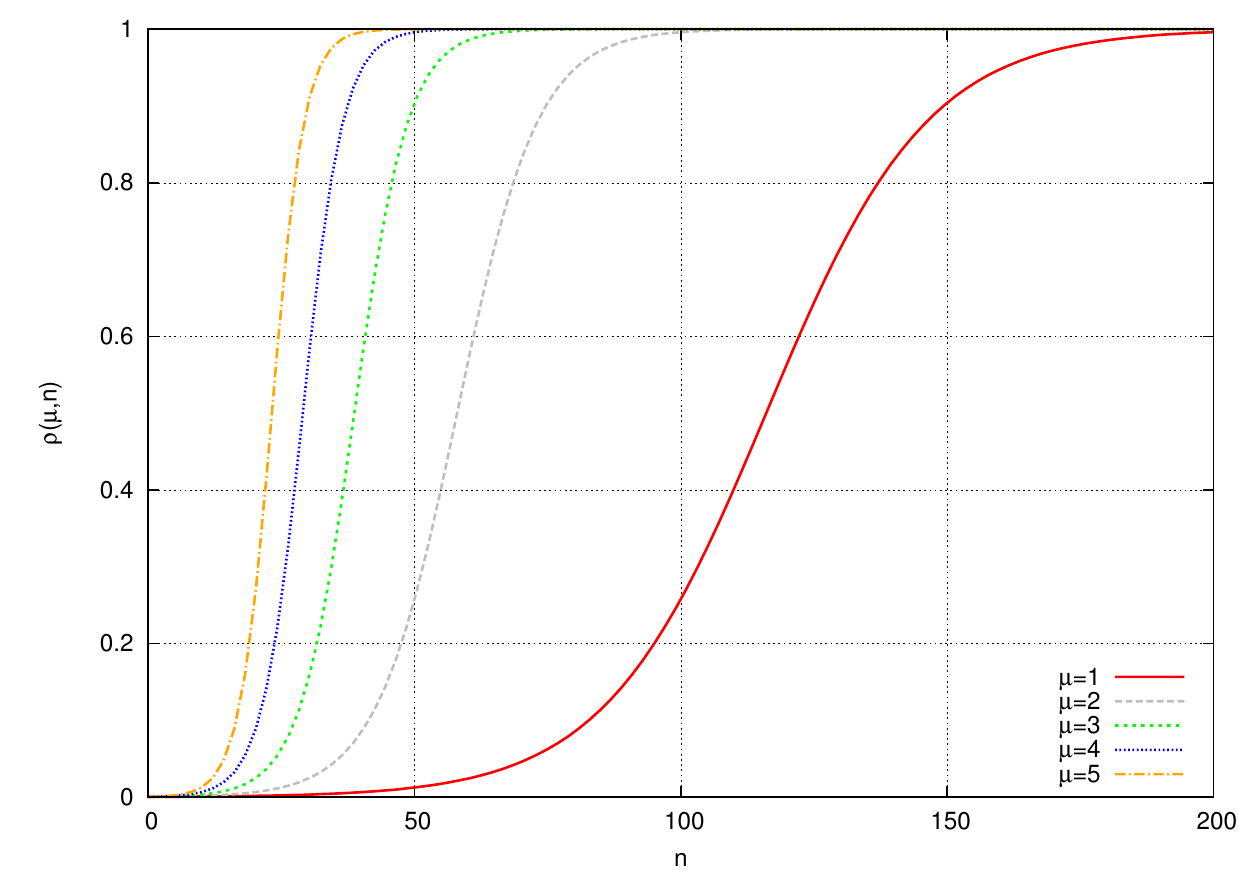}
	\caption{We plotted Eq. \eqref{eqn:popularityScore} for five different values of $\mu$, in order to show the corresponding change in the shape of the logistic curve. Clearly, for a low average rating $\mu$ a large number of ratings is necessary to achieve a high popularity score.}\label{im:rhoPlot}
\end{figure}

We calculate a popularity score, $\rho$, for each item based on the number of ratings received during the critical period and the average rating at the end of the critical period. In order to achieve a score that ranges between 0 and 1, a logistic function is used. A logistic function is an s-shaped curve that is frequently used to model population growth \citep{Pearl1920}. The function grows exponentially first with the slope decreasing thereafter until the function reaches a steady state. The popularity score $\rho$ is given by:

\begin{equation}\label{eqn:popularityScore}
\rho(\mu,n) = \frac{1}{1+e^{-k(\mu n-c)}},
\end{equation} 

where $\mu$ is the average rating, $n$ is the number of ratings received within the critical period and $c$ and $k$ are constants. The constant $c$ is chosen as follows: In the MovieLens dataset for instance, a movie is considered as popular if it received 29 ratings or more and obtained an average score of 4 or higher. The constant $c$ is chosen such that a movie that received exactly 29 ratings and an average score of 4 receives a popularity score of $\rho=0.5$. The constant $k$ is chosen such that an item without any ratings receives a popularity score of approximately equal to 0. See Section S2 in the supplementary material for calculations of $c$ and $k$ for the different datasets. Figure \ref{im:rhoPlot} shows the change in the shape of $\rho(\mu,n)$ as the average rating $\mu$ changes in the case of the MovieLens dataset. Figure S3 in the supplementary material shows a plot of $\rho(\mu,n)$ in the form of a heat map to give a clear visualisation. 

Some rating systems may not give their users the opportunity to rate an item on a scale. One example is the Digg network, where an edge between a user and a story indicates that the user liked the story. To be able to calculate the popularity score $\rho$ for items in such networks, we assign a rating of 5 to each edge.

\begin{figure}[t]
	\centering
	\includegraphics[width=0.8\textwidth]{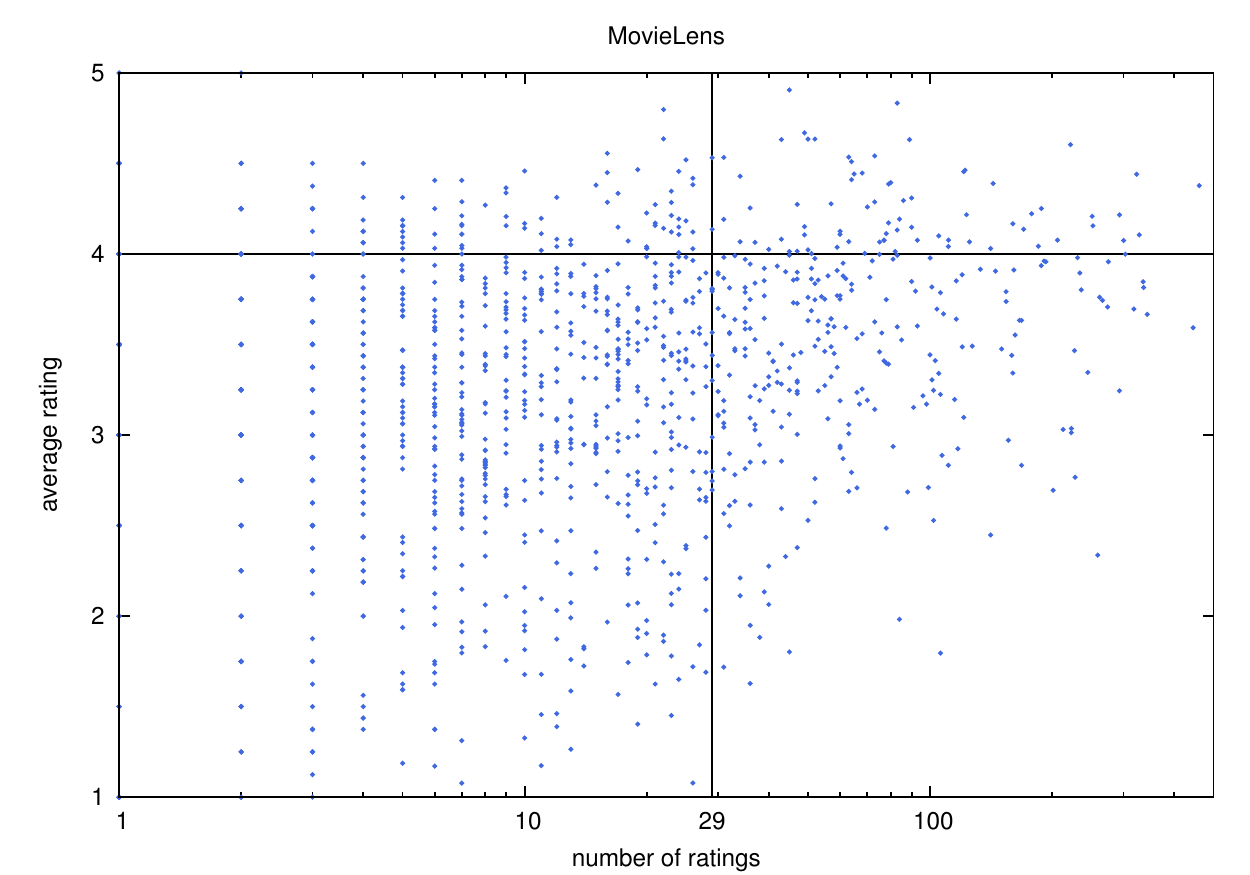}
	\caption{We plotted the movies' average ratings $\mu$ against the number of ratings $n$ they received, to examine their relationship. The correlation coefficient between the average rating and the number of ratings is low (0.183). If we were to consider only the number of received ratings to determine the movie's popularity, many would be wrongly classified as popular, see lower right quadrant of the plot. All movies in the lower right quadrant received a relatively high number of ratings but have an average rating of less than 4.}\label{im:numberVsAverage}	
\end{figure}

In case of the MovieLens network, we expect that the average rating of an item is to some degree related to the number of ratings it received. However, Fig. \ref{im:numberVsAverage} shows that just using the number of ratings to predict the average rating results in  a large number of items (see lower right quadrant of the plot) being wrongly identified as popular.

Since new items are hard to classify as popular or unpopular, we examine the user who is the first to rate the new item. We start by extracting the ego network of the user who first rated a new item, to depth three. In other words, we include all first, second and third neighbours of the ego and only allow edges corresponding to ratings made during a certain period of time prior to the first rating of the new item. Analysis of user activity showed that in the case of the MovieLens network, ratings made up to ten days prior to the first rating of the new movie have to be included. Any period less than ten days, in most cases, resulted in an ego network that only contained a single edge. Since the dynamics in the Digg network is much faster, a period of six hours prior to the first rating is sufficient. Since we consider the ego network of the first user, we henceforth refer to this user as the ego. A depth of three is necessary to be able to calculate the local clustering coefficients of the ego.

Since the popularity score $\rho$ of an item is dependent on both the number of ratings received as well as the average rating, the two parameters are separately predicted. The popularity scores thus obtained are compared to the actual popularity scores calculated from the real data.

\subsection{Predicting the number of ratings}\label{subsec:numberRatings}

As will be demonstrated below, both the ego's degree, i.e. the ego's rating activity, as well as the number of its second neighbours perform poorly as predictors, whereas, the ego's clustering behaviour is a better predictor of the number of ratings that the new item will receive.

\subsubsection{The Ego's Rating Activity}\label{subsubsec:degree}
In the context of one-mode networks, it has been argued many times that a node with high degree is not necessarily influential \citep[e.g.][]{Li2014}. This is also the case in the MovieLens and Digg networks. In other words, the rating of a new item by a highly active user does not imply that the item will receive many ratings. Figure S4 in the supplementary material demonstrates this.

\subsubsection{Second Neighbours of the Ego}\label{subsubsec:secondNeighbours}
Since rating networks are bipartite, it would be more apt to consider the number of second neighbours of an ego as a predictor of an item's number of ratings instead of the ego's degree, as the latter only gives the the number of items rated by the ego. Second neighbours of the ego are users who rated at least one item that was also rated by the ego. 

In addition, users who rated the same items as the ego in the recent past are more likely to rate the new item than a randomly selected user. Hence, the number of second neighbours of the ego, may give some indication of the number of ratings that the new item will receive in the near future. 

However, as depicted in Fig. S5 in the supplementary material, the number of second neighbours is also a poor indicator for the item's future degree. 

\subsubsection{The ego's clustering behaviour}\label{subsubsec:clusteringBehaviour}
We now examine the clustering behaviour of users who were the first to rate a new item. Table \ref{tab:egoNets} shows that the extracted ego networks vary considerably.

\begin{table}[h]
	\centering
	\begin{tabular}{llll}
		\toprule
		& & MovieLens & Digg\\
		\midrule
		& range & [3, 4411] & [25, 11769]\\
		size & mean & 1681 & 1521\\
		& sd & 1060 & 1504\\
		\midrule
		& range & [2, 540]& [2, 38]\\
		mean degree & mean & 82& 8\\
		& sd & 69& 4\\
		\midrule
		& range& [0.0098, 1]& [0.0038, 0.5217]\\
		density & mean & 0.1065& 0.0397\\
		& sd & 0.1793& 0.0477\\
		\bottomrule
	\end{tabular}
	\caption{The table shows the range, mean and standard deviation of the size, average degree and density of the extracted ego networks.}\label{tab:egoNets}
\end{table}

To determine the ego's clustering behaviour, we calculate the four different local clustering coefficients that we introduced in Section \ref{sec:clustering}. For the majority of the ego networks, the induced clustering coefficient is the highest, whereas in most cases the three other coefficients are 0. Clustering in all of the ego networks is low due to their scale-free nature. 

The four different clustering coefficients measure the proportions of induced 6-cycles, 6-cycles with one chord, 6-cycles with two chords and 6-cycles with three chords respectively. In the case of one-mode networks \citet{Ugander2012} have shown that a node is less likely to be infected if its neighbourhood is well connected.  Hence, we expect that when an ego's clustering coefficient is lower than the average clustering coefficient in its ego network, then the new item will receive a high number of ratings in the near future. \citet{Chen2013} have also shown that nodes with a low local clustering coefficient attract more connections.

We compare the ego's clustering behaviour to all other users in its ego network by calculating how many standard deviations it lies away from the average local clustering coefficient over all users in the ego network. We expect that a high difference in standard deviations indicates that the new item will receive many ratings in the future, provided the ego's clustering coefficient is lower than the average. If on the other hand, the ego's clustering coefficient is higher than the average and the difference in standard deviations is high, we expect the new item to receive very few ratings in the future. 

Since the three chord clustering coefficient, $icc_{(3)}$ shows higher connectivity than the induced clustering coefficient $icc_{(0)}$, we give the differences in standard deviations appropriate weights, according to their level of connectivity. The first rating of the new item is also taken into account, since it is likely to influence other users. For instance, if the first rating of a new item is low, it is less likely to receive many ratings than if the first rating is high. 

The following equation gives the predicted number of ratings $\hat{n}$ if the ego's clustering coefficients are lower than the average:

\begin{equation}\label{eqn:prediction}
\hat{n} = \frac{r}{3}(2{\Delta icc_{(ego,0)}} + 3{\Delta icc_{(ego,1)}} + 4{\Delta icc_{(ego,2)}} + 5{\Delta icc_{(ego,3)}}),
\end{equation}

where $r$ is the first rating of the new item that was given by the ego and $\Delta icc_{(ego,k)}$ is the difference in standard deviations between that particular clustering coefficient of the ego and the average of the same clustering coefficient in the ego's network. Since ratings range between 1 and 5, the initial rating $r$ is divided by 3. Hence, a rating of 3 is treated as neutral.

If, on the other hand, one or more of the ego's clustering coefficients are higher than the average, then we divide by the corresponding weight instead of multiplying. For example, if in a given ego network, $icc_{(ego,0)}$ and $icc_{(ego,1)}$ are lower that the corresponding average clustering coefficient and the other two local clustering coefficients $icc_{(ego,2)}$ and $icc_{(ego,3)}$ are higher, then Eq. \eqref{eqn:prediction} becomes: $\hat{n} = \frac{r}{3}(2\Delta icc_{(ego,0)} + 3\Delta icc_{(ego,1)} + \Delta icc_{(ego,2)}/4 + \Delta icc_{(ego,3)}/5)$. 

The reason for only using the first reviewer of an item to predict its popularity is that our aim is to make predictions as early as possible. Considering a combination of the first few ratings may improve predictions, however, it comes with the disadvantage of having to make the predictions later in time. Whether the improvement of predictions is high enough to sacrifice early predictions will be answered in future work.

Figure S6 in the supplementary material shows that the number of ratings was correctly predicted for 53\% of movies in the MovieLens network and 67\% of stories in the Digg network. We took a closer look at the movies where the number of ratings was incorrectly predicted. Generally, the movies that received a higher number of ratings than predicted, received very mixed reviews by critics. We have listed some of these movies in Table S1 in the supplementary material. Among the movies that received less ratings than predicted, are non-English films. Usually these movies received good critique in their country of origin. We give a more detailed discussion in Section S4 of the supplementary material.

\subsection{Predicting the Average Rating}
We again make use of the logistic curve to estimate the average rating of items. Since ratings range between 1 and 5, the function values should also range between these values and hence

\begin{equation}\label{eqn:ratings}
f(n) =  1+\frac{4}{1+e^{-k(n-c)}},
\end{equation}

where $c$ and $k$ are constants. The constants are chosen such that $f(n) \approx 1$ if an item has a predicted number of ratings equal to zero and $f(n) = 4$ if an item has a predicted number of ratings equal to the number of ratings that the average item received. Calculations are shown in the supplementary material.

Using Eq. \eqref{eqn:ratings} together with the first rating $r$ and the predicted number of ratings $\hat{n}$, we can estimate the future average rating, $\hat{\mu}$, of the new movie:

\begin{equation}\label{eqn:predRating}
\hat{\mu} = (f(\hat{n}) + r)/2.
\end{equation}

Equation \eqref{eqn:ratings} predicts the future average rating well. However, the rating that is given by the first user has a large influence on other users. Therefore, we take the average between $f(\hat{n})$ and the first rating to improve prediction of the future average rating of an item.

With the two parameters, $\hat{\mu}$ and $\hat{n}$, the popularity score can now be predicted for each item. 

\begin{figure}[h]
	\centering
	\includegraphics[width=1.0\textwidth]{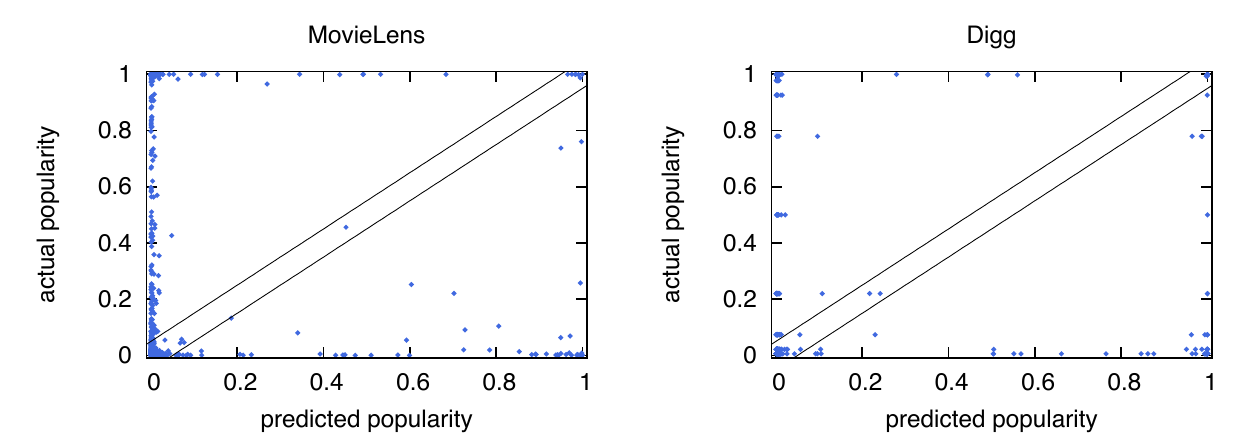}
	\caption{We plotted the actual popularity of items in the MovieLens (left) and Digg (right) networks as a function of the predicted popularity. For the MovieLens network we were able to correctly predict the future popularity of 66\% of the movies. For the Digg network we achieved a success rate of 51\%. }\label{im:predictions}
\end{figure}

We have predicted the popularity of all movies that were released between 2004 and 2007 in the MovieLens network and 350 randomly chosen news stories in the Digg network. Figure \ref{im:predictions} compares the predicted popularity scores $\hat{\rho}$ (Eq. \eqref{eqn:popularityScore}) to the actual popularity scores $\rho$ and shows that our method correctly predicted the popularity of 638 of the 962 movies in the MovieLens network, with the difference between the predicted popularity $\hat{\rho}$ and the actual popularity $\rho$ being less than 0.05. This is a success rate of approximately 66\%. This is especially good since no use was made of any other information about the new movie other than the first rating. In previous research, authors were only able to predict the popularity of a new movie with 30\% accuracy \citep{Zeng2013}. In the Digg network we achieved a success rate of approximately 51\%, where we were able to correctly predict the popularity of 179 out of 350 news stories, compared to a 20\% success rate achieved in \citep{Zeng2013}. 

\section{Conclusion}
In this paper, we introduced a new clustering coefficient that is suitable for the analysis of rating networks as well as other bipartite networks that form in a similar manner. Since our clustering coefficient distinguishes between differently structured clusters it gives valuable topological information about the network. 

Further, we showed that it may be used as a tool to predict the future popularity of items in rating networks. We focused on improving predictions of new items that are hard to classify as popular or unpopular due to lack of information. Previous research of one-mode networks has shown that nodes with a low clustering coefficient attract connections in the future \citep{Chen2013a}. These nodes are also known to spread information faster than nodes with a higher clustering coefficient. \citet{Ugander2012} have also shown that a node is less likely to be infected if its neighbourhood is well connected. We consider the clustering coefficient as a predictor of popularity based on this previous research. Since a new item, at the time of prediction, has degree one, we examine the clustering coefficient of the user who rated the item first. If this user has a low clustering coefficient compared to the users in its neighbourhood, the missing edges may be added by building connections to the new item.

By investigation of the clustering behaviour of the user who made the first rating of a new item we correctly predicted the future popularity of over 65\% of new movies in the MovieLens dataset and over 50\% of new stories in the Digg dataset. This is a major improvement on previous research. 

One limitation of the clustering coefficient is that computation time grows exponentially with the size of the network. Thus, a topic for future research would be the improvement of algorithms that count the different 5-paths and 6-cycles. 

Another very interesting topic of research is the identification of the, for instance 100, most popular items in the future. We aim to address this problem in future work. 

\clearpage
\appendix
\section{Supplementary Figures}

\begin{figure}[h]
	\centering
	\includegraphics[width=1.0\textwidth]{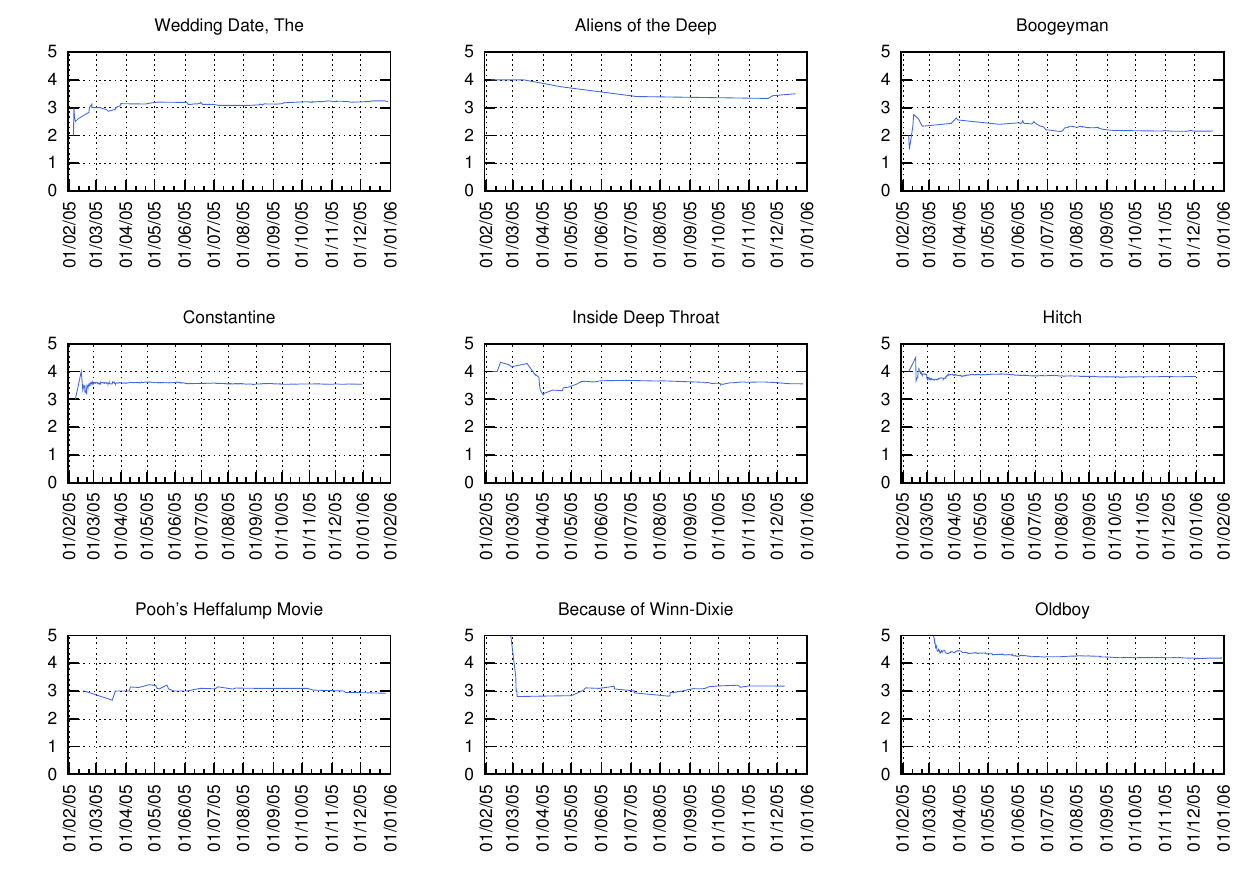}
	\caption{In the MovieLens network it is often the case that the average rating of a movie fluctuates during the first month after the initial rating, before becoming steady. Here, we have plotted nine examples. The $x$-axis shows the time in form of the date and the $y$-axis displays the corresponding average rating. }
\end{figure}

\begin{figure}[h]
	\centering
	\includegraphics[width=1.0\textwidth]{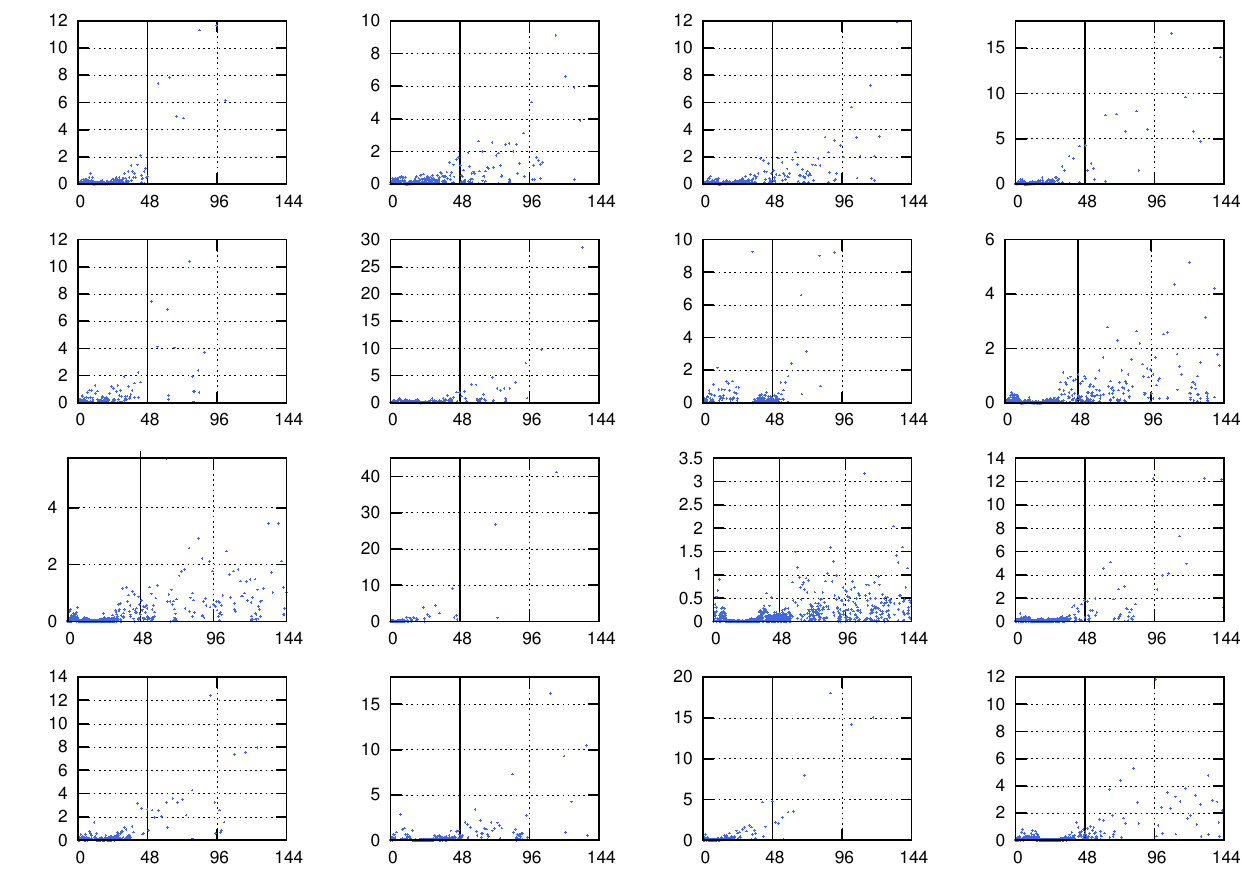}
	\caption{In the Digg network the user interest in an item decays very quickly. Here we plotted the differences in time between the ratings of 16 different stories. The $y$-axis shows the difference in hours to the previous rating and the $x$-axis shows the time in hours after the first rating. Most items are frequently rated during the first 48 hours after the initial rating.}
\end{figure}

\begin{figure}[h]
	\includegraphics[width=1.0\textwidth]{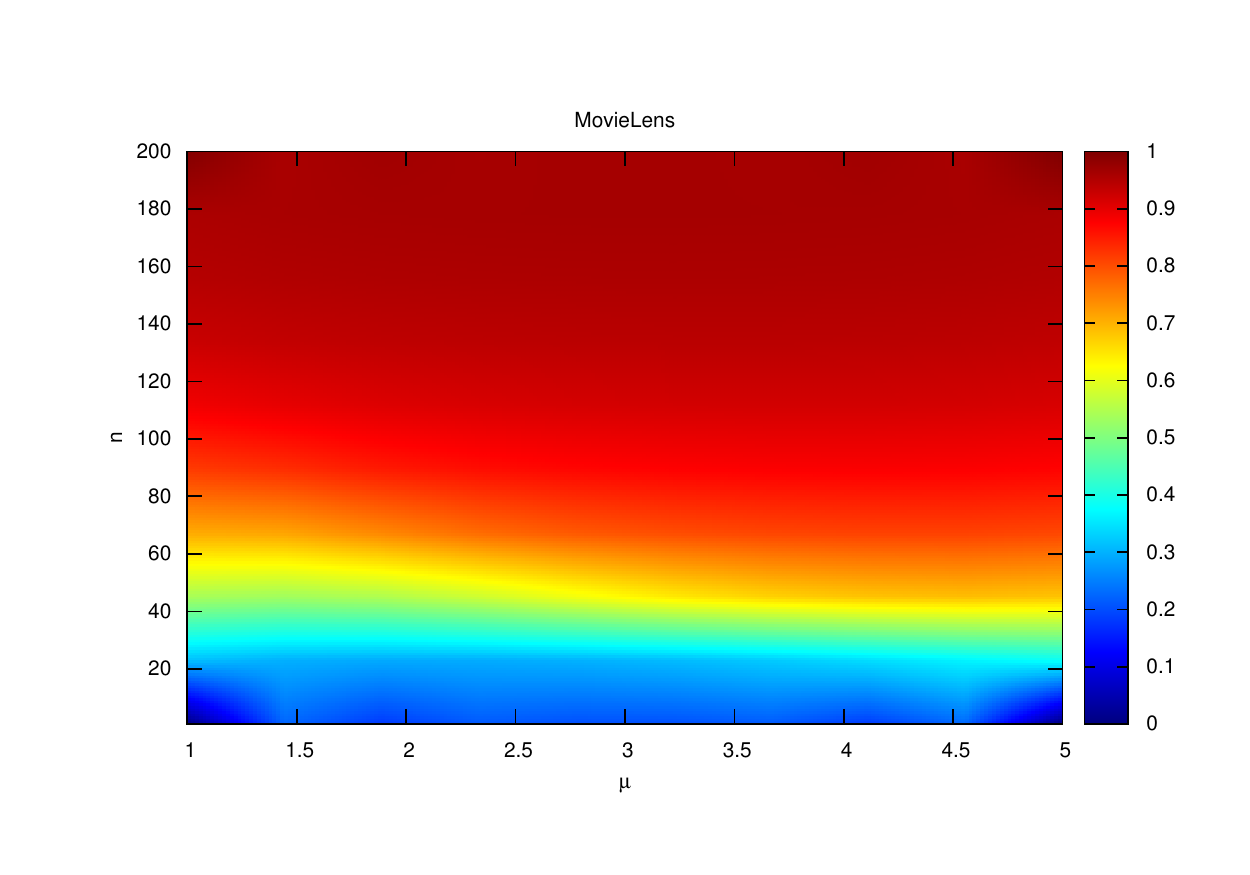}
	\caption{The heat map shows values of the popularity score $\rho$ (Eq. (6)) with respect to the number of ratings and the average rating for the MovieLens network. Red corresponds to $\rho=1$ and blue to $\rho\approx 0$. The heat map shows that a high value of $\rho$ can only be achieved if the number of ratings as well as the average rating is high. The upper left corner of the heat map is also dark red, however, an item with a low average rating usually does not gain the requisite large number of ratings in order to receive a high popularity score. }
\end{figure}	

\begin{figure}[h]
	\includegraphics[width=1.0\textwidth]{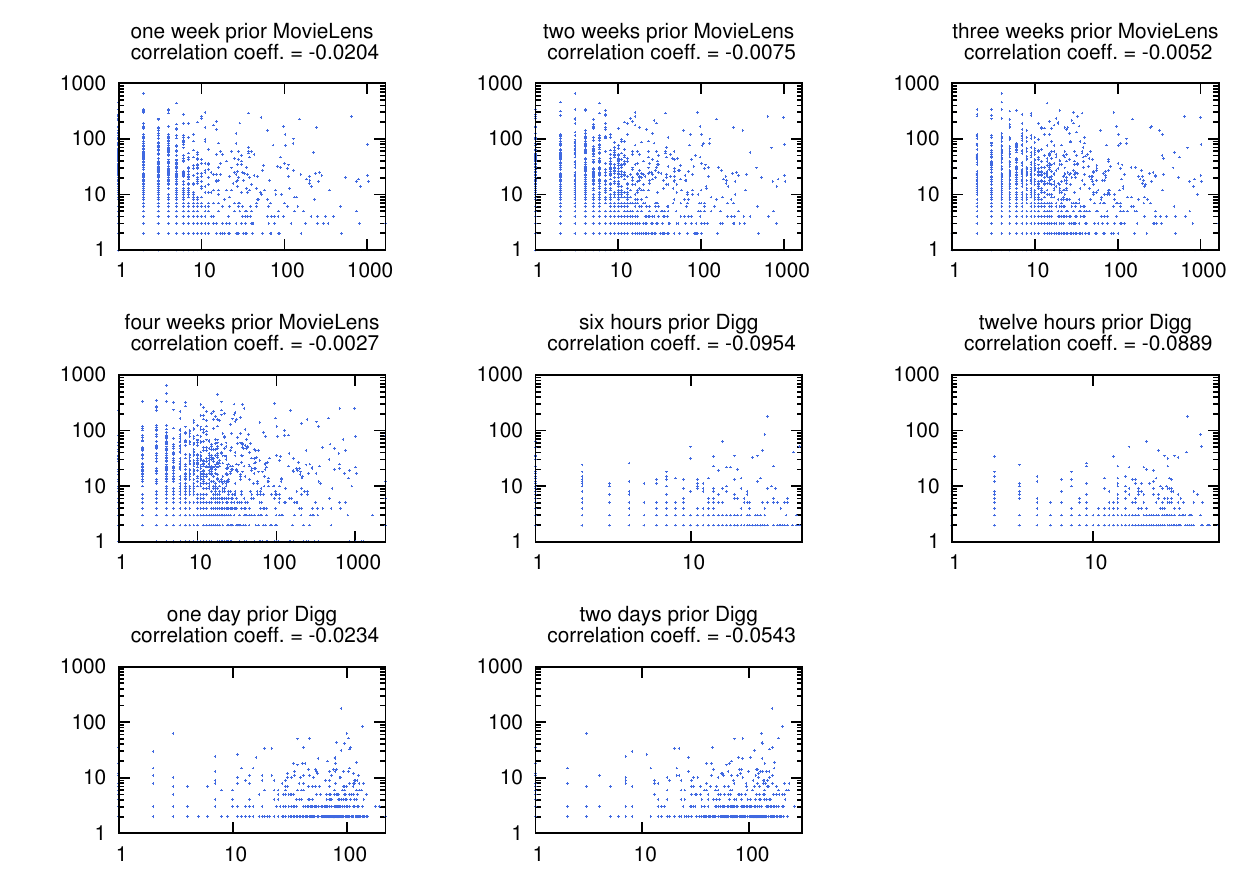}
	\caption{In order to demonstrate that the ego's degree is not a good indicator for how many ratings the new item will receive, we have plotted the new items' degrees ($y$-axis) after the critical period against the corresponding ego's degree ($x$-axis). The first four plots correspond to the MovieLens network, the last four plots correspond to the Digg network. We considered ego degrees one, two, three and four weeks prior to the first rating in case of MovieLens. For the Digg network we considered ego degrees six hours, twelve hours, one day and two days prior to the first rating. In all cases no correlation is present (see title of each plot).}
\end{figure}

\begin{figure}[h]
	\includegraphics[width=1.0\textwidth]{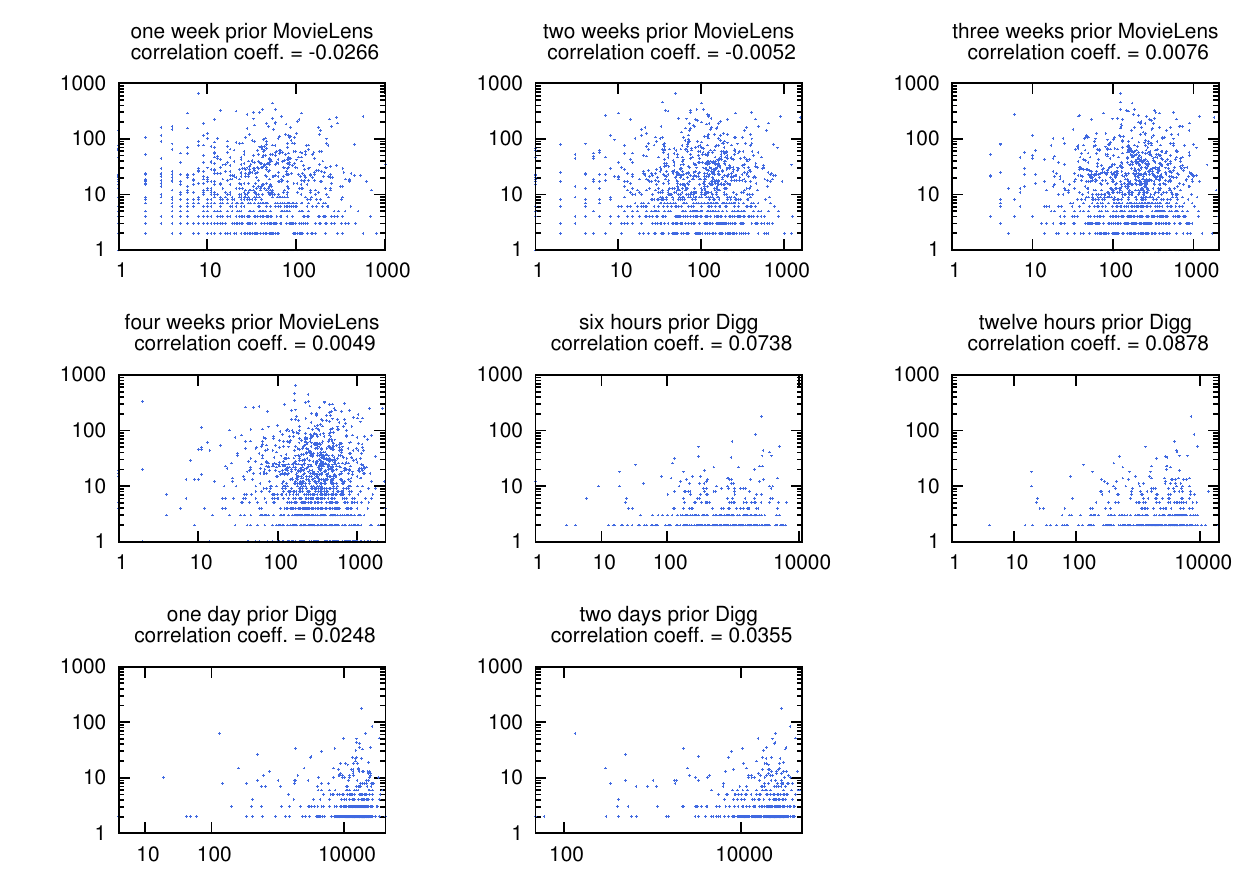}
	\caption{In order to demonstrate that the ego's number of second neighbours is not a good indicator for how many ratings the new item will receive, we have plotted the new items' degrees ($y$-axis) after the critical period against the corresponding number of second neighbours of the ego ($x$-axis). The first four plots correspond to the MovieLens network, the last four plots correspond to the Digg network. We considered ego degrees one, two, three and four weeks prior to the first rating in case of MovieLens. For the Digg network we considered ego degrees six hours, twelve hours, one day and two days prior to the first rating. In all cases no correlation is present (see title of each plot).}
\end{figure}

\begin{figure}[h]
	\includegraphics[width=1.0\textwidth]{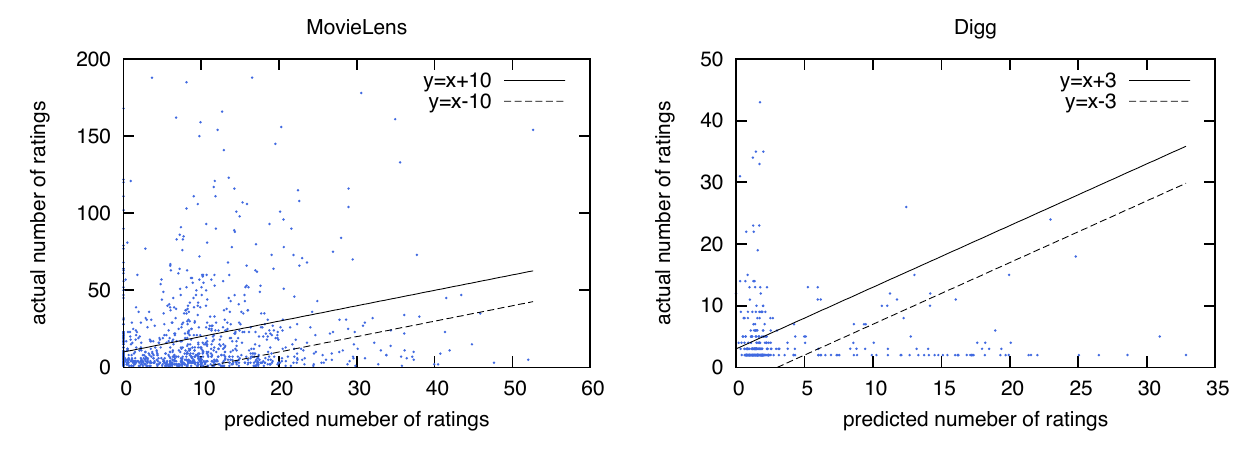}
	\caption{We plotted the actual number of received ratings, $n$, as a function of the predicted number of ratings, $\hat{n}$. For the MovieLens network (left), we correctly predicted the number of ratings for 53\% of the movies. These movies are represented by dots within the two straight lines. For the Digg network (right), we correctly predicted the number of ratings for 57\% of news stories. We give a detailed analysis of the movies in the MovieLens network where the number of ratings was incorrectly predicted in Section \ref{incorrectlyPredicted}.}
\end{figure}

\clearpage
\section{Calculation of the Constants in Eq. (6)}
The popularity score $\rho(\mu,n)$ is given by: $\rho(\mu,n) = 1/(1+e^{-k(\mu n-c)})$, see Eq. (6), where $\mu$ is the average rating, $n$ is the number of ratings received within the critical period and $c$ and $k$ are constants. 

The constants $c$ and $k$ are chosen such that an item that received as many rating as the average item and an average rating of 4 receives a popularity score of 0.5. An item without any ratings should receive a popularity score of approximately equal to 0. 

Hence, in case of the MovieLens network:

\begin{eqnarray*}
	0.5 &=& 1/(1+e^{-k(4*29-c)})\\
	0.5(1+e^{-k(116-c)}) &=& 1\\
	e^{-k(116-c)} &=& 1\\
	-k(116-c) &=& 0 \\
	c &=& 116 \quad \textrm{since}\quad k\neq 0
\end{eqnarray*}

and

\begin{eqnarray*}
	0.5\cdot10^{-3} &=& 1/(1+e^{-k(0-116)})\\
	0.5\cdot10^{-3}(1+e^{116k}) &=& 1\\
	e^{116k} &=& 1,999\\
	116k &=& 7.6 \\
	k &=& 0.066.
\end{eqnarray*}

Therefore $\rho(\mu,n) = 1/(1+e^{-0.066(\mu n-116)})$ in the MovieLens network. 

In the Digg network a story received on average six ratings during the first two days. Since we associated every edge in the Digg network with a rating of 5, a news story that received six ratings and an average rating of 5 should receive a popularity score of 0.5. Hence, $c=30$, $k=0.253$ and therefore $\rho(\mu,n) = 1/(1+e^{-0.253(\mu n-30)})$ for the Digg network.

\section{Calculation of the Constants in Eq. (8)}
The predicted average of a new item is given by $f(n) =  1+ 4/(1+e^{-k(n-c)})$, where $c$ and $k$ are constants, see Eq. (8). The constants are chosen such that $f(n)\approx 1$ if an item has a predicted number of ratings equal to zero and $f(n) = 4$ if an item has a predicted number of ratings equal to the number of ratings that the average item received. In the MovieLens network the average item received 29 ratings and hence,

\begin{eqnarray*}
	0.5\cdot10^{-3} &=& 4/(1+e^{-k(0-c)})\\
	0.5\cdot10^{-3}(1+e^{ck}) &=& 4\\
	e^{ck} &=& 7999\\
	ck &=& 8.987 \\
	c &=& 8.987/k
\end{eqnarray*}

and 

\begin{eqnarray*}
	3 &=& 4/(1+e^{-k(29-c)})\\
	3(1+e^{-k(29-c)}) &=& 4\\
	e^{-k(29-c)} &=& 1/3\\
	-29k +ck &=& -1.099 \\
	29k &=& 10.086\\
	k &=& 0.348.
\end{eqnarray*}

Therefore, $f(n) =  1+ 4/(1+e^{-0.348(n-25.825)})$ for the MovieLens network. In case of the Digg network, the average rating does not need to be predicted, since we associated every edge with a rating of 5.

\section{Analysis of Movies}\label{incorrectlyPredicted}
We have predicted the popularity of all movies that were released between 2004 and 2007 in the MovieLens network. Our method correctly predicted the number of ratings of 510 of the 962 movies. This is a success rate of approximately 53\%. Here we take a close look at the movies where the number of ratings was incorrectly predicted. We divide this set of movies into two subsets, one containing movies where the actual number of ratings is higher than predicted and the other containing movies where the actual number of ratings is lower than predicted. 

Amongst the set of movies where the actual number of ratings was lower than predicted, approximately 30\% of movies are in languages other than English. These movies generally received very positive reviews by film critics. Hence our predictions overall agree with film critics and the low actual number of ratings in the MovieLens dataset may be explained by the high number of English speaking users. Table \ref{tab:nonEnglinshMovies} lists some of these movies together with the number of ratings that were collected by the websites Rotten Tomatoes (http://www.rottentomatoes.com/) and Metacritic (http://www.metacritic.com/). In addition, for the website Rotten Tomatoes the table displays the tomatometer score that represents the percentage of approved critics that have given the movie a positive review. For Metacritic we also show the metascore. The metascore ranges between 0 and 100, with 100 being the best possible score.

\begin{center}
	\begin{longtable}{p{5cm} p{1.5cm} p{1.5cm} p{2cm} p{1.5cm}}
		\toprule
		Movie title & actual number of ratings (average) & predicted number of ratings (predicted average) & Rotten Tomatoes (average) & Metacritic (average)\\
		\midrule
		\endhead
		Red Lights (Feux rouges) (2004) & 3 (2.88) & 19 (2.17) & 86 (83\%) & 28 (74)\\
		Lost Embrace (El Abrazo Partido) (2004) & 2 (3) & 17 (1.59) & 48 (83\%) & 23 (70)\\
		Sea Inside, The (Mar adentro) (2004) & 7 (4.11) & 19 (2.66) & 131 (84\%) & 38 (74)\\
		Machuca (2004) & 7 (2.28) & 21 (3.08) & 33 (89\%) & 37 (76)\\
		Tae Guk Gi - The Brotherhood of War (Taegukgi hwinalrimyeo) (2004) & 5 (4.06) & 37 (4.22) & 41 (80\%) & 19 (64)\\
		Appleseed (Appurush\={\i}do) (2004) & 7 (4.21) & 22 (2.91) & 32 (25\%) & 17 (40)\\
		Turtles Can Fly (Lakposhth\^{a} ham parvaz mikonand) (2004) & 2 (3.5) & 12 (2.02) & 72 (88\%) & 31 (85)\\
		Loop the Loop (a.k.a. Up and Down) (Horem p\'{a}dem) (2004) & 2 (4.25) & 40 (4.49) & 65 (83\%) & 27 (78)\\
		Walk on Water (2004) & 3 (4.25) & 18 (3.13) & 75 (72\%) & 28 (65)\\
		Look at Me (Comme une image) (2004) & 5 (3.31)& 27 (4.17) & 98 (87\%) & 30 (79)\\
		Year of the Yao, The (2004) & 1 (4) & 17 (2.6) & 33 (67\%) & 11 (62)\\
		Three... Extremes (Saam gaang yi) (2004) & 8 (3.84) & 34 (4.38) & 62 (84\%) & 22 (66)\\
		Bittersweet Life, A (Dalkomhan insaeng) (2005) & 4 (3.25) & 17 (2.35) & 10 (100\%) & na\\
		Duck Season (Temporada de patos) (2004) & 5 (3.38) & 19 (2.92) & 73 (90\%) & 27 (74)\\
		Usphizin (2004) & 7 (3.91) & 17 (2.85) & 61 (93\%) & na\\
		Vinci (2004) & 2 (3.5) & 29 (3.97) & na & na\\
		Business, The (2005) & 3 (3.63) & 20 (2.48) & na & na\\
		Tony Takitani (2004) & 3 (3.38) & 22 (2.65) & 52 (88\%) & 22 (88)\\
		Child, The (L'Enfant) (2005) & 6 (3.94) & 20 (3.26) & na & 34 (87)\\
		Hidden Blade, The (Kakushi ken oni no tsume) (2004) & 4 (3.75) & 29 (4.01) & 31 (87\%) & 11 (76)\\
		Three Times (Zui Hao De Shi Guang) (2005) & 4 (3.06) & 19 (2.68) & 50 (86\%) & 22 (80)\\
		Taxidermia (2006) & 5 (4.16)& 19 (2.4) & 46 (80\%) & 9 (83)\\
		Gui Si (Silk) (2006) & 1 (3.5)& 12 (2.26) & 5 (40\%) & na\\
		Arn - The Knight Templar (Arn - Tempelriddaren) (2007) & 3 (3.5) & 16 (2.56) & na & na\\
		Tell No One (Ne le dis \`{a}  personne) (2007) & 8 (3.71)& 34 (4.88) & 108 (94\%) & 30 (82)\\
		Czech Dream (\v{C}esk\'{y} sen) (2004) & 2 (3.75) & 19 (2.4) & 24 (79\%) & 7 (72)\\
		4 Months, 3 Weeks and 2 Days (4 luni, 3 s\v{a}pt\v{a}m\^{a}ni \c{s}i 2 zile) (2007) & 35 (3.82) & 46 (4.5) & 133 (95\%) & 37 (97)\\
		Om Shanti Om (2007) & 3 (3.88) & 25 (2.94) & 13 (77\%) & na\\
		Aerial, The (La Antena) (2007) & 3 (4) & 48 (4.99) & 11 (64\%) & na\\
		Inside (\`{A} l'int\'{e}rieur) (2007) & 5 (3.13) & 17 (2.35) & 12 (83\%) & na \\
		Unknown Solider, The (Unbekannte Soldat, Der) (2006) & 1 (3) & 13 (2.02) & 10 (60\%) & 6 (71)\\
		Aleksandra (2007) & 1 (3) & 13 (2.02) & na & 13 (85)\\
		Ganes (2007) & 2 (3) & 20 (2.22) & na & na\\
		Katyn (2007) & 4 (3.63) & 21 (1.8) & 64 (94\%) & 17 (81)\\
		Maria Full of Grace (Maria, Llena eres de gracia) (2004) & 7 (4) & 19 (2.44) & 139 (97\%) & 39 (87)\\
		Veer Zaara (2004) & 2 (3.25) & 40 (4.99) & na & 5 (67)\\
		Bad Education (La Mala educaci\'{o}n) (2004) & 15 (3.79) & 30 (4.58) & 137 (88\%) & 34 (81)\\
		\bottomrule
		\caption{The table lists the non-English movies that were predicted to receive a higher than the actual number of ratings. In general, these movies received very positive reviews from critics. The low number of ratings received by MovieLens users may be due to their demographics. We listed the number of ratings that were recorded by the websites Rotten Tomatoes and Metacritic.  In addition, for the website Rotten Tomatoes the table displays the tomatometer score that represents the percentage of approved critics that have given the movie a positive review. For Metacritic we also show the metascore. The metascore ranges between 0 and 100, with 100 being the best possible score.}\label{tab:nonEnglinshMovies}
	\end{longtable}
\end{center}

Among the movies that our method predicted to receive a lower number of ratings than in reality are many that received very mixed or negative reviews from other websites. Since we do not have any information about the MovieLens users, we are unable to explain these results. It may be possible that in these cases, the user who first rated the movie usually does not watch movies in that particular genre. Another reason may be that these movies were highly anticipated and therefore received many ratings, although scores were generally low. 

\section*{Acknowledgment}
The authors would like to thank the members of the GroupLens Research Project \citep{Grouplens2014} and the authors of KONECT \citep{Konect2014} for providing the data on their website. They would also like to thank the anonymous reviewers for their constructive comments that greatly helped in improving the paper. 

\bibliography{mybibfile}

\begin{thebibliography}{33}
\expandafter\ifx\csname natexlab\endcsname\relax\def\natexlab#1{#1}\fi
\providecommand{\url}[1]{\texttt{#1}}
\providecommand{\href}[2]{#2}
\providecommand{\path}[1]{#1}
\providecommand{\DOIprefix}{doi:}
\providecommand{\ArXivprefix}{arXiv:}
\providecommand{\URLprefix}{URL: }
\providecommand{\Pubmedprefix}{pmid:}
\providecommand{\doi}[1]{\href{http://dx.doi.org/#1}{\path{#1}}}
\providecommand{\Pubmed}[1]{\href{pmid:#1}{\path{#1}}}
\providecommand{\bibinfo}[2]{#2}
\ifx\xfnm\relax \def\xfnm[#1]{\unskip,\space#1}\fi
\bibitem[{Aral \& Walker(2012)}]{Aral2012}
\bibinfo{author}{Aral, S.}, \& \bibinfo{author}{Walker, D.}
  (\bibinfo{year}{2012}).
\newblock \bibinfo{title}{{Identifying influential and susceptible members of
  social networks}}.
\newblock {\it \bibinfo{journal}{Science}\/},  {\it \bibinfo{volume}{337}\/}.
\bibitem[{Asratian et~al.(1998)Asratian, Denley \&
  H\"{a}ggkvist}]{Asratian1998}
\bibinfo{author}{Asratian, A.~S.}, \bibinfo{author}{Denley, T.~M.}, \&
  \bibinfo{author}{H\"{a}ggkvist, R.} (\bibinfo{year}{1998}).
\newblock {\it \bibinfo{title}{{Bipartite Graphs and Their Applications}}\/}.
\newblock (\bibinfo{edition}{1st} ed.).
\newblock \bibinfo{address}{Cambridge}: \bibinfo{publisher}{Cambridge
  University Press}.
\bibitem[{Barabasi \& Albert(1999)}]{Barabasi1999}
\bibinfo{author}{Barabasi, A.-L.}, \& \bibinfo{author}{Albert, R.}
  (\bibinfo{year}{1999}).
\newblock \bibinfo{title}{{Emergence of scaling in random networks}}.
\newblock {\it \bibinfo{journal}{Science}\/},  {\it \bibinfo{volume}{286}\/},
  \bibinfo{pages}{11}.
\bibitem[{Chen et~al.(2013{\natexlab{a}})Chen, Gao, L\"{u} \& Zhou}]{Chen2013a}
\bibinfo{author}{Chen, D.-B.}, \bibinfo{author}{Gao, H.},
  \bibinfo{author}{L\"{u}, L.}, \& \bibinfo{author}{Zhou, T.}
  (\bibinfo{year}{2013}{\natexlab{a}}).
\newblock \bibinfo{title}{{Identifying influential nodes in large-scale
  directed networks: the role of clustering.}}
\newblock {\it \bibinfo{journal}{PloS one}\/},  {\it \bibinfo{volume}{8}\/},
  \bibinfo{pages}{e77455}.
\bibitem[{Chen et~al.(2013{\natexlab{b}})Chen, Xiao, Zeng \& Zhang}]{Chen2013}
\bibinfo{author}{Chen, D.-B.}, \bibinfo{author}{Xiao, R.},
  \bibinfo{author}{Zeng, A.}, \& \bibinfo{author}{Zhang, Y.-C.}
  (\bibinfo{year}{2013}{\natexlab{b}}).
\newblock \bibinfo{title}{{Path diversity improves the identification of
  influential spreaders}}.
\newblock {\it \bibinfo{journal}{Europhys. Lett.}\/},  {\it
  \bibinfo{volume}{104}\/}, \bibinfo{pages}{68006}.
\bibitem[{Conaldi et~al.(2012)Conaldi, Lomi \& Tonellato}]{Conaldi2012}
\bibinfo{author}{Conaldi, G.}, \bibinfo{author}{Lomi, A.}, \&
  \bibinfo{author}{Tonellato, M.} (\bibinfo{year}{2012}).
\newblock \bibinfo{title}{{Dynamic models of affiliation and the network
  structure of problem solving in an open source software project}}.
\newblock {\it \bibinfo{journal}{Organizational Res. Methods}\/},  {\it
  \bibinfo{volume}{15}\/}, \bibinfo{pages}{385--412}.
\bibitem[{{de Solla Price}(1976)}]{Price1976}
\bibinfo{author}{{de Solla Price}, D.} (\bibinfo{year}{1976}).
\newblock \bibinfo{title}{{A general theory of bibliometric and other
  cumulative advantage processes}}.
\newblock {\it \bibinfo{journal}{J. of the Am. Soc. for Inf. Sci.}\/},  {\it
  \bibinfo{volume}{27}\/}, \bibinfo{pages}{292--306}.
\bibitem[{Diestel(2005)}]{Diestel2005}
\bibinfo{author}{Diestel, R.} (\bibinfo{year}{2005}).
\newblock {\it \bibinfo{title}{{Graph Theory}}\/}.
\newblock (\bibinfo{edition}{3rd} ed.).
\newblock \bibinfo{address}{New York}: \bibinfo{publisher}{Springer}.
\bibitem[{GoupLens(2014)}]{Grouplens2014}
\bibinfo{author}{GoupLens} (\bibinfo{year}{2014}).
\newblock \bibinfo{title}{{MovieLens 10M}}.
\newblock \URLprefix \url{http://grouplens.org/datasets/movielens/}.
\bibitem[{Kitsak et~al.(2010)Kitsak, Gallos, Havlin, Liljeros, Muchnik, Stanley
  \& Makse}]{Kitsak2010}
\bibinfo{author}{Kitsak, M.}, \bibinfo{author}{Gallos, L.~K.},
  \bibinfo{author}{Havlin, S.}, \bibinfo{author}{Liljeros, F.},
  \bibinfo{author}{Muchnik, L.}, \bibinfo{author}{Stanley, H.~E.}, \&
  \bibinfo{author}{Makse, H.~A.} (\bibinfo{year}{2010}).
\newblock \bibinfo{title}{{Identifying influential spreaders in complex
  networks}}.
\newblock {\it \bibinfo{journal}{Nature Phys.}\/},  {\it
  \bibinfo{volume}{6}\/}, \bibinfo{pages}{36}.
\bibitem[{{KONECT}(2014)}]{Konect2014}
\bibinfo{author}{{KONECT}} (\bibinfo{year}{2014}).
\newblock \bibinfo{title}{{Digg votes}}.
\newblock \URLprefix \url{http://konect.uni-koblenz.de/networks/digg-votes}.
\bibitem[{Li et~al.(2014)Li, Zhou, L\"{u} \& Chen}]{Li2014}
\bibinfo{author}{Li, Q.}, \bibinfo{author}{Zhou, T.}, \bibinfo{author}{L\"{u},
  L.}, \& \bibinfo{author}{Chen, D.} (\bibinfo{year}{2014}).
\newblock \bibinfo{title}{{Identifying influential spreaders by weighted
  LeaderRank}}.
\newblock {\it \bibinfo{journal}{Physica A}\/},  {\it \bibinfo{volume}{404}\/},
  \bibinfo{pages}{47--55}.
\bibitem[{Liebig \& Rao(2014)}]{Liebig2014}
\bibinfo{author}{Liebig, J.}, \& \bibinfo{author}{Rao, A.}
  (\bibinfo{year}{2014}).
\newblock \bibinfo{title}{{Identifying influential nodes in bipartite networks
  using the clustering coefficient}}.
\newblock In {\it \bibinfo{booktitle}{Proceedings of the 2014 Tenth
  International Conference on Signal-Image Technology and Internet-Based
  Systems}\/} (pp. \bibinfo{pages}{323--330}).
\bibitem[{Lind et~al.(2005)Lind, Gonz\~{a}lez \& Herrmann}]{Lind2005}
\bibinfo{author}{Lind, P.~G.}, \bibinfo{author}{Gonz\~{a}lez, M.~C.}, \&
  \bibinfo{author}{Herrmann, H.~J.} (\bibinfo{year}{2005}).
\newblock \bibinfo{title}{{Cycles and clustering in bipartite networks}}.
\newblock {\it \bibinfo{journal}{Phys. Rev. E}\/},  {\it
  \bibinfo{volume}{72}\/}, \bibinfo{pages}{056127}.
\bibitem[{Liu et~al.(2013)Liu, Ren \& Guo}]{Liu2013}
\bibinfo{author}{Liu, J.~G.}, \bibinfo{author}{Ren, Z.~M.}, \&
  \bibinfo{author}{Guo, Q.} (\bibinfo{year}{2013}).
\newblock \bibinfo{title}{{Ranking the spreading influence in complex
  networks}}.
\newblock {\it \bibinfo{journal}{Physica A}\/},  {\it \bibinfo{volume}{392}\/},
  \bibinfo{pages}{4154--4159}.
\bibitem[{L\"{u} et~al.(2011)L\"{u}, Zhang, Yeung \& Zhou}]{lu2011}
\bibinfo{author}{L\"{u}, L.}, \bibinfo{author}{Zhang, Y.~C.},
  \bibinfo{author}{Yeung, C.~H.}, \& \bibinfo{author}{Zhou, T.}
  (\bibinfo{year}{2011}).
\newblock \bibinfo{title}{{Leaders in social networks, the Delicious case}}.
\newblock {\it \bibinfo{journal}{PLoS ONE}\/},  {\it \bibinfo{volume}{6}\/},
  \bibinfo{pages}{e21202}.
\bibitem[{Medo et~al.(2011)Medo, Cimini \& Gualdi}]{Medo2011}
\bibinfo{author}{Medo, M.}, \bibinfo{author}{Cimini, G.}, \&
  \bibinfo{author}{Gualdi, S.} (\bibinfo{year}{2011}).
\newblock \bibinfo{title}{{Temporal effects in the growth of networks}}.
\newblock {\it \bibinfo{journal}{Phys. Rev. Lett.}\/},  {\it
  \bibinfo{volume}{107}\/}, \bibinfo{pages}{238701}.
\bibitem[{Newman(2010)}]{Newman2010}
\bibinfo{author}{Newman, M. E.~J.} (\bibinfo{year}{2010}).
\newblock {\it \bibinfo{title}{{Networks: An Introduction}}\/}.
\newblock (\bibinfo{edition}{1st} ed.).
\newblock \bibinfo{address}{Oxford}: \bibinfo{publisher}{Oxford University
  Press}.
\bibitem[{Opsahl(2013)}]{Opsahl2013}
\bibinfo{author}{Opsahl, T.} (\bibinfo{year}{2013}).
\newblock \bibinfo{title}{{Triadic closure in two-mode networks: Redefining the
  global and local clustering coefficients}}.
\newblock {\it \bibinfo{journal}{Soc. Netw.}\/},  {\it \bibinfo{volume}{35}\/},
  \bibinfo{pages}{159--167}.
\bibitem[{Page et~al.(1999)Page, Brin, Motwani \& Winograd}]{Page1999}
\bibinfo{author}{Page, L.}, \bibinfo{author}{Brin, S.},
  \bibinfo{author}{Motwani, R.}, \& \bibinfo{author}{Winograd, T.}
  (\bibinfo{year}{1999}).
\newblock \bibinfo{title}{{The PageRank citation ranking: Bringing order to the
  web}}.
\newblock {\it \bibinfo{journal}{Stanf. InfoLab}\/}, .
\bibitem[{Pearl \& Reed(1920)}]{Pearl1920}
\bibinfo{author}{Pearl, R.}, \& \bibinfo{author}{Reed, L.~J.}
  (\bibinfo{year}{1920}).
\newblock \bibinfo{title}{{On the rate of growth of the population of the
  United States since 1790 and its mathematical representation}}.
\newblock {\it \bibinfo{journal}{Proceedings of the National Academy of
  Sciences}\/},  {\it \bibinfo{volume}{6}\/}, \bibinfo{pages}{275--288}.
\bibitem[{Ren et~al.(2014)Ren, Zeng, Chen, Liao \& Liu}]{Ren2014}
\bibinfo{author}{Ren, Z.-M.}, \bibinfo{author}{Zeng, A.},
  \bibinfo{author}{Chen, D.-B.}, \bibinfo{author}{Liao, H.}, \&
  \bibinfo{author}{Liu, J.-G.} (\bibinfo{year}{2014}).
\newblock \bibinfo{title}{{Iterative resource allocation for ranking spreaders
  in complex networks}}.
\newblock {\it \bibinfo{journal}{Europhys. Lett.}\/},  {\it
  \bibinfo{volume}{106}\/}, \bibinfo{pages}{48005}.
\bibitem[{Resnick et~al.(1994)Resnick, Iacovou, Suchak, Bergstrom \&
  Riedl}]{Resnick1994}
\bibinfo{author}{Resnick, P.}, \bibinfo{author}{Iacovou, N.},
  \bibinfo{author}{Suchak, M.}, \bibinfo{author}{Bergstrom, P.}, \&
  \bibinfo{author}{Riedl, J.} (\bibinfo{year}{1994}).
\newblock \bibinfo{title}{{GroupLens : An open architecture for collaborative
  filtering of netnews}}.
\newblock In {\it \bibinfo{booktitle}{Proceedings of the 1994 ACM Conference on
  Computer Supported Cooperative Work}\/} (pp. \bibinfo{pages}{175--186}).
\bibitem[{Robins \& Alexander(2004)}]{Robins2004}
\bibinfo{author}{Robins, G.~L.}, \& \bibinfo{author}{Alexander, M.}
  (\bibinfo{year}{2004}).
\newblock \bibinfo{title}{{Small worlds among interlocking directors: Network
  structure and distance in bipartite graphs}}.
\newblock {\it \bibinfo{journal}{Computational \& Mathematical Organization
  Theory}\/},  {\it \bibinfo{volume}{10}\/}, \bibinfo{pages}{69--94}.
\bibitem[{Schafer et~al.(2007)Schafer, Frankowski, Herlocker \&
  Sen}]{Schafer2007}
\bibinfo{author}{Schafer, J.}, \bibinfo{author}{Frankowski, D.},
  \bibinfo{author}{Herlocker, J.}, \& \bibinfo{author}{Sen, S.}
  (\bibinfo{year}{2007}).
\newblock \bibinfo{title}{{Collaborative filtering recommender systems}}.
\newblock In {\it \bibinfo{booktitle}{The Adaptive Web}\/} (pp.
  \bibinfo{pages}{291--324}).
\bibitem[{Ugander et~al.(2012)Ugander, Backstrom, Marlow \&
  Kleinberg}]{Ugander2012}
\bibinfo{author}{Ugander, J.}, \bibinfo{author}{Backstrom, L.},
  \bibinfo{author}{Marlow, C.}, \& \bibinfo{author}{Kleinberg, J.}
  (\bibinfo{year}{2012}).
\newblock \bibinfo{title}{{Structural diversity in social contagion.}}
\newblock {\it \bibinfo{journal}{Proceedings of the National Academy of
  Sciences of the United States of America}\/},  {\it \bibinfo{volume}{109}\/},
  \bibinfo{pages}{5962--6}.
\bibitem[{Vogt \& Mestres(2010)}]{Vogt2010}
\bibinfo{author}{Vogt, I.}, \& \bibinfo{author}{Mestres, J.}
  (\bibinfo{year}{2010}).
\newblock \bibinfo{title}{{Drug-target networks}}.
\newblock {\it \bibinfo{journal}{Molecular Inform.}\/},  {\it
  \bibinfo{volume}{29}\/}, \bibinfo{pages}{10--14}.
\bibitem[{Wei et~al.(2013)Wei, Deng, Zhang, Deng \& Mahadevan}]{Wei2013}
\bibinfo{author}{Wei, D.}, \bibinfo{author}{Deng, X.}, \bibinfo{author}{Zhang,
  X.}, \bibinfo{author}{Deng, Y.}, \& \bibinfo{author}{Mahadevan, S.}
  (\bibinfo{year}{2013}).
\newblock \bibinfo{title}{{Identifying influential nodes in weighted networks
  based on evidence theory}}.
\newblock {\it \bibinfo{journal}{Physica A}\/},  {\it \bibinfo{volume}{392}\/},
  \bibinfo{pages}{2564--2575}.
\bibitem[{Zeng et~al.(2013)Zeng, Gualdi, Medo \& Zhang}]{Zeng2013}
\bibinfo{author}{Zeng, A.}, \bibinfo{author}{Gualdi, S.},
  \bibinfo{author}{Medo, M.}, \& \bibinfo{author}{Zhang, Y.-C.}
  (\bibinfo{year}{2013}).
\newblock \bibinfo{title}{{Trend prediction in temporal bipartite networks: The
  case of MovieLens, Netflix, and Digg}}.
\newblock {\it \bibinfo{journal}{Advances in Complex Syst.}\/},  {\it
  \bibinfo{volume}{16}\/}, \bibinfo{pages}{1350024}.
\bibitem[{Zeng \& Zhang(2013)}]{Zeng2013a}
\bibinfo{author}{Zeng, A.}, \& \bibinfo{author}{Zhang, C.~J.}
  (\bibinfo{year}{2013}).
\newblock \bibinfo{title}{{Ranking spreaders by decomposing complex networks}}.
\newblock {\it \bibinfo{journal}{Phys. Lett. A}\/},  {\it
  \bibinfo{volume}{377}\/}, \bibinfo{pages}{1031--1035}.
\bibitem[{Zhang et~al.(2008)Zhang, Wang, Li, Li, Di \& Fan}]{Zhang2008}
\bibinfo{author}{Zhang, P.}, \bibinfo{author}{Wang, J.}, \bibinfo{author}{Li,
  X.}, \bibinfo{author}{Li, M.}, \bibinfo{author}{Di, Z.}, \&
  \bibinfo{author}{Fan, Y.} (\bibinfo{year}{2008}).
\newblock \bibinfo{title}{{Clustering coefficient and community structure of
  bipartite networks}}.
\newblock {\it \bibinfo{journal}{Physica A}\/},  {\it \bibinfo{volume}{387}\/},
  \bibinfo{pages}{6869--6875}.
\bibitem[{Zhang et~al.(2013)Zhang, Zhu, Wang \& Zhao}]{Zhang2013}
\bibinfo{author}{Zhang, X.}, \bibinfo{author}{Zhu, J.}, \bibinfo{author}{Wang,
  Q.}, \& \bibinfo{author}{Zhao, H.} (\bibinfo{year}{2013}).
\newblock \bibinfo{title}{{Identifying influential nodes in complex networks
  with community structure}}.
\newblock {\it \bibinfo{journal}{Knowl.-Based Syst.}\/},  {\it
  \bibinfo{volume}{42}\/}, \bibinfo{pages}{74--84}.
\bibitem[{Zhou et~al.(2007)Zhou, Ren, Medo \& Zhang}]{Zhou2007}
\bibinfo{author}{Zhou, T.}, \bibinfo{author}{Ren, J.}, \bibinfo{author}{Medo,
  M.}, \& \bibinfo{author}{Zhang, Y.~C.} (\bibinfo{year}{2007}).
\newblock \bibinfo{title}{{Bipartite network projection and personal
  recommendation}}.
\newblock {\it \bibinfo{journal}{Phys. Rev. E}\/},  {\it
  \bibinfo{volume}{76}\/}, \bibinfo{pages}{046115}.

\end{thebibliography}

\end{document}